\documentclass{aa}

\usepackage{graphicx}
\usepackage{txfonts}
\usepackage{natbib}
\bibpunct{(}{)}{;}{a}{}{,} 
\usepackage{aalongtable,lscape}

\newcommand{\m}{\,$\mu$m}
\newcommand{\spitzer}{{\it Spitzer}}
\newcommand{\HII}{\ion{H}{ii}}
\newcommand{\FeII}{[\ion{Fe}{ii}]}
\newcommand{\NeII}{[\ion{Ne}{ii}]}
\newcommand{\NeIII}{[\ion{Ne}{iii}]}
\newcommand{\ArII}{[\ion{Ar}{ii}]}
\newcommand{\ArIII}{[\ion{Ar}{iii}]}
\newcommand{\SIII}{[\ion{S}{iii}]}
\newcommand{\SIV}{[\ion{S}{iv}]}
\newcommand{\lsun}{$L_{\odot}$}
\newcommand{\msun}{$M_{\odot}$}
\newcommand{\brg}{Br-$\gamma$}
\newcommand{\pab}{Pa-$\beta$}
\newcommand{\hal}{H-$\alpha$}

\begin{document}
   \title{High Resolution IR Observations of the Starburst Ring in
     NGC\,7552}

   \subtitle{One Ring to Rule Them All?}

   \author{B.~R. Brandl    
	  \inst{1}
          \and
          N.\,L. Mart\'{\i}n-Hern\'{a}ndez
          \inst{2,3}
          \and
          D. Schaerer
          \inst{4,5}
          \and
          M. Rosenberg
          \inst{1}
          \and
          P.~P. van der Werf
          \inst{1}
          }

   \offprints{B.~R. Brandl, brandl@strw.leidenuniv.nl}

   \institute{Leiden Observatory, Leiden University, P.O. Box 9513, 
             NL-2300 RA Leiden, Netherlands
             \and
             Instituto de Astrof\'isica de Canarias, V\'ia L\'actea s/n,
             E-38205 La Laguna, Spain
             \and
             Departamento de Astrof\'isica, Universidad de La Laguna, 
             E-38205 La Laguna, Tenerife, Spain
             \and
	     Geneva Observatory, University of Geneva, 51, Ch. des Maillettes, 
             CH-1290 Versoix, Switzerland
             \and
             CNRS, IRAP, 9 Av. colonel Roche, BP 44346, F-31028 Toulouse 
             cedex 4, France
             }

   \date{Received June 26, 2011; accepted May 4, 2012}

\abstract{} 
{Approximately 20\% of all spiral galaxies display starburst activity
  in nuclear rings of a few hundred parsecs in diameter.  It is our
  main aim to investigate how the starburst ignites and propagates
  within the ring, leading to the formation of massive stellar
  clusters.}
{We observed the ring galaxy NGC\,7552 with the mid-infrared (MIR)
  instrument VISIR at an angular resolution of 0\farcs$3 - 0$\farcs4
  and with the near-infrared (NIR) integral-field spectrograph SINFONI
  on the VLT, and complement these observations with data from ISO and
  Spitzer.}
{The starburst ring is clearly detected at MIR wavelengths at the
  location of the dust-extincted, dark ring seen in HST observations.
  This ``ring'', however, is a rather complex annular region of more
  than 100 parsec width.  We find a large fraction of diffuse
  \NeII\ and PAH emission in the central region that is not associated
  with the MIR peaks on spatial scales of $\sim30$\,pc. We do not
  detect MIR emission from the nucleus of NGC\,7552, which is very
  prominent at optical and NIR continuum wavelengths.  However, we
  have identified nine unresolved MIR peaks within the ring. The
  average extinction of these peaks is $A_{\rm V} = 7.4$ and their
  total infrared luminosity is $L_{IR} = 2.1\times 10^{10}$\lsun.  The
  properties of these peaks are typical for MIR-selected massive
  clusters found in other galaxies.  The ages of the MIR-selected
  clusters are in the range of $5.9\pm0.3$~Myr.  The age spread among
  the clusters of 0.8\,Myr is small compared to the travel time of
  $\sim5.6$\,Myr for half an orbit within the starburst ring.  We find
  no strong evidence for a scenario where the continuous inflow of gas
  leads to the ongoing formation of massive clusters at the contact
  points between galactic bar and starburst ring. Instead, it appears
  more likely that the gas density build up more gradually over larger
  ring segments, and that the local physical conditions govern cluster
  formation.  We note that the fundamental limitation on the accurate
  derivation of cluster age, mass and IMF slope is the lack of higher
  angular resolution.  Resolving the highly embedded, massive clusters
  requires milli-arcsecond resolution at infrared wavelengths, which
  will be provided by the next generation of instruments on extremely
  large telescopes (ELTs).
}
  {}

  
  \titlerunning{The Starburst Ring in NGC\,7552}

   \maketitle
%

\section{Introduction}
\label{sect:intro}

Massive star formation in galaxies occurs in various appearances and
modes, from localised {\sc H II} regions and blue compact dwarf
galaxies to luminous starbursts in ultra-luminous infrared galaxies
(ULIRGs) and sub-millimeter galaxies.  One particularly interesting
starburst mode shows a ring-like morphology around the galactic
nucleus \citep[e.g.,][]{sarzi07,boker08,mazzuca08}.  These systems are
not uncommon -- in approximately 20\% of all spiral galaxies
star-formation occurs primarily in these rings of typically a few
hundred parsecs in diameter \citep{knapen05}.  These large structures
of intense star formation are primarily found in barred spiral
galaxies of types Sa-Sbc.  In fact, bars are expected to account for
about 3.5~times more triggered central star formation than galaxy
interactions \citep{ellison11}.  Strong magnetic fields
  \citep[$105\mu$G in the ring of NGC\,7552, ][]{beck11} are likely to
  create magnetic stress that causes inflow of gas toward the center.

The key question is under what conditions the large amounts of gas and
dust, necessary to form the massive young star clusters, accumulate
and concentrate in these ring-like structures.  Unfortunately, the
situation is not clear, partially due to the lack of high resolution
observations at longer wavelengths and partially due to the complexity
of the dynamics of gas and dust, and hence several scenarios have been
discussed in the literature.

According to \citet{shlosman90}, the torque from the large-scale bar
weakens near its centre where the influence of the bulge begins to
dominate the disk potential. The gas concentrates in ring-like
resonances where the gas is free of gravitational torques.  In
contrast, \citet{kenney93} explain the rings as a consequence of a
previous nuclear starburst which has used up the gas in the centre,
leaving the nuclear ring as a remnant.  Recently, \citet{vaisanen11}
reported on an outward propagating ring of star formation in NGC\,1614
resulting from a relatively major merger event.

Arguably the most common explanation is that the gas accumulates
during the inflow around the radii at which the stellar orbits
experience dynamical resonances with the rotating bar potential
\citep[e.g.,][]{binney08}. In the central region this typically
happens at the so-called inner Lindblad resonance (ILR), arising from
the interplay between the bar and the stellar orbits
\citep[e.g.,][]{telesco93,buta96,boker08}.  The gas is trapped between
an inner ILR and an outer ILR.  If the potential allows only one ILR,
the gas will concentrate directly at the ILR, if the system has two
ILRs, the gas will accumulate at the inner ILR -- \citep[see review by
][ for more details]{buta96}.

However, more recently \citet{regan03} claimed that the common
assumption that nuclear rings are related to an ILR is incorrect, and
showed in simulations that there is no resonance at the inner Lindblad
resonance in barred galaxies.  They argue that confusion has developed
between the orbit family transition that occurs at these radii in weak
bars and a true resonance.  Furthermore, \citet{regan03} found that
the radius of nuclear rings decreases with time, either because the
rings accumulate lower specific angular momentum gas or because of
dissipation at the contact point of the bar dust lane and the nuclear
ring.

\begin{figure}[ht]
    \centering
    \resizebox{\hsize}{!}{\includegraphics{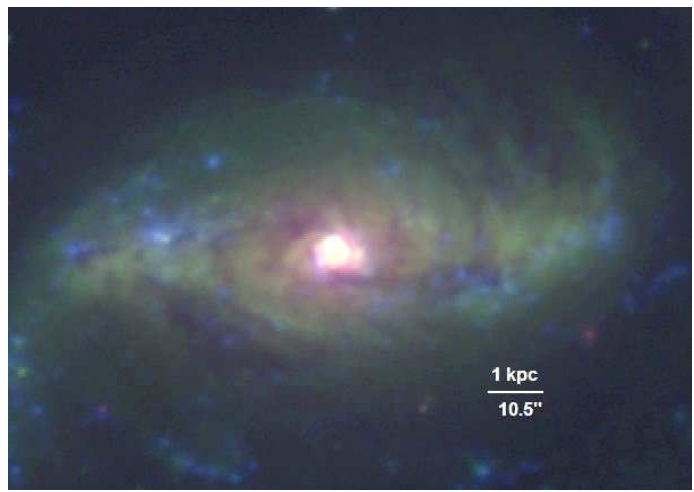}}
    \caption{Colour composite of NGC 7552 made from images provided by
      the SINGS team \citep{kennicutt03} in \hal\ (blue), and the B
      and I bands (green and red, respectively), taken on the CTIO 1.5
      meter telescope.  North is up and East to the left.  A scale bar
      indicating angular and absolute distances is also shown.}
    \label{fig:wherebar}
  \end{figure}

Obviously, observations may be able to provide further insights.  A
very good example of a starburst ring galaxy is NGC\,7552, a nearly
face-on, barred spiral galaxy at an inclination of
$\sim28$\degr\ \citep{feinstein90}.  Originally classified as
amorphous \citep{sersic65} it appears in the RC3 catalogue
\citep{deVaucouleurs91} as morphological type SB(s)ab and in the RSA
catalogue \citep{sandage81} as type SBbc(s).  Both designations
indicate a barred spiral (SB) galaxy with spiral arms that spring from
the ends of a bar (s), and an outer ring (R').
Fig.~\ref{fig:wherebar} shows a 3-color image of NGC\,7552 to
illustrate the global morphology and the relevant scales.  Throughout
this paper we assume a distance of 19.5~Mpc \citep{tully88}; at this
distance, 1\arcsec\ corresponds to 95~pc.

NGC\,7552 has been previously classified as a LINER galaxy, based on
the [O\,I]$\lambda$6300 line \citep{durret88}.  Neither X-ray
\citep{liu05} nor NIR observations \citep{forbes94b} did reveal
significant activity from a galactic nucleus.  In fact, models of gas
flow in barred galaxies \citep{piner95} have shown that the gas flow,
necessary to fuel an active galactic nucleus, can be interrupted by a
nuclear ring.  The global spectral energy distribution (SED), from UV
to radio, is well characterised \citep[e.g.,][]{dale07}.  From the
IRAS colours of the entire galaxy, \citet{sanders03} estimated a total
infrared luminosity of $8.5\times 10^{10}$\lsun\ (adjusted for the
assumed distance).

Several previous studies \citep[e.g.,][]{feinstein90, forbes94,
  schinnerer97, siebenmorgen04, vanderwerf06} provided strong evidence
for a starburst ring in the centre of NGC\,7552
(section~\ref{starburstring}).  While the macroscopic picture of how
these rings form seems well established, the morphology of the central
region and the microscopic picture of how, when and where the
starburst ignites and propagates are rather uncertain and require more
and better observational constraints.  It is the main aim of this
paper to investigate these details.

MIR observations provide excellent diagnostics to detect embedded
clusters\footnote{Throughout this paper we use the physical term
  `cluster' and the observable `MIR peak' as synonyms.  Whether or not
  this is correct will be discussed in
  section~\ref{sect:clusterpeak}.} \citep[e.g.][]{gorjian01,
  beck02,vacca02} and quantify massive star formation
\citep[e.g.][]{genzel00, peeters04}.  Previous studies of NGC\,7552 at
MIR wavelengths have been severly limited in spatial
resolution, with angular resolutions of $0.8'' - 1''$
\citep{schinnerer97, siebenmorgen04} and $\sim4''$ \citep{dale07}.
The availability of the VLT Spectrometer and Imager for the
Mid-Infrared (VISIR) at the ESO VLT, makes it possible to study
galaxies in the N~band (8 -- 13\m) at an angular resolution of
0\farcs3 under excellent seeing conditions\footnote{A seeing of
  0\farcs3 at 10\m\ corresponds to 0\farcs55 optical seeing.
  0\farcs3 is also the diffraction-limited resolution of an 8\,m
  telescope at 10\m.}.

In this paper we complement our mid-IR data with NIR images at similar
angular resolution.  The latter data were taken with SINFONI, a NIR
integral-field spectrograph at the ESO VLT.  Combining near- and
mid-IR photometry will provide essential information on both, the
stellar sources that provide most of the energy output, and their
dusty, re-radiating environment.  The observations presented here,
with their unprecedented high angular resolution, will allow us to
study the morphology of the starburst ring, the distribution and
properties of its star clusters, and their formation history.

This paper is structured as follows. Section~\ref{obs} describes the
observations and data reduction. Section~\ref{analysis} presents the
photometric and spectroscopic results, and sect.~\ref{discussion}
provides a detailed discussion of our results.  Finally,
sect.~\ref{summary} summarises our main conclusions.


\section{Observations and Data Reduction}
\label{obs}


\subsection{VISIR}


\subsubsection{Imaging}

The imaging data were acquired with the VLT Imager and Spectrograph
for the mid-IR \citep[VISIR, ][]{lagage04,pantin05}, mounted at the
Cassegrain focus of the VLT Unit Telescope 3, Melipal.  The data were
obtained in period of March 2005 to January 2006 in nine different
filters.  Table~\ref{table:fluxes} lists the relevant filters, their
central wavelengths and spectral widths.  A cross-comparison with the
spectro-photometry (section~\ref{secspectroscopy}) shows overall very
good agreement in eight narrow and broad-band filters, but with
  the exception of the SIV\_1 filter, in which the measured flux
  density appears to be about 60\% too high (see Fig.~\ref{fig:sed}).
  However, the $9.73 - 9.91\mu$m SIV\_1 filter has intrinsically very
  low transmission, resulting in a four times lower nominal
  sensitivity with respect to comparable VISIR
  filters\footnote{http://www.eso.org/sci/facilities/paranal/instruments/visir/inst/},
  and was thus not included in the analysis.

\begin{table*}
\caption{\label{table:fluxes}  
The spectral properties and the measured flux densities in units of
mJy through the nine used VISIR filters.  In addition to the
individual MIR peaks M1 through M9 we list their summed contribution
($\Sigma$M$_{\rm i}$), the integrated fluxes within an annulus of
1\farcs5\,$\le r \le$\,4\arcsec, and the image background noise
($\sigma$ in units of mJy\,arcsec$^{-2}$). Upper limits are defined as
three times the image noise within a circle of 0\farcs7 in diameter.} 
\begin{center}
\begin{tabular}{l@{\hspace{7pt}}c@{\hspace{7pt}}c@{\hspace{7pt}}c@{\hspace{7pt}}c@{\hspace{7pt}}c@{\hspace{7pt}}c@{\hspace{7pt}}c@{\hspace{7pt}}c@{\hspace{7pt}}c}
\hline\hline\\[-5pt] Filter & SIC & PAH1 & ArIII & SIV\_1 & SIV & SIV\_2 & NeII\_1 & NeII & NeII\_2\\ \hline\\[-5pt]
$\lambda_c\ [\mu\rm m]$ & 11.85 & 8.59 & 8.99 & 9.82 & 10.49 & 10.77 & 12.27 & 12.80 & 13.03 \\ 
$\Delta\lambda\ [\mu$m] &  2.34 & 0.49 & 0.14 & 0.18 &  0.16 &  0.19 &  0.18 &  0.21 &  0.22 \\ \hline\\[-5pt]
M1  & 116 $\pm$  13  &  67 $\pm$  10  &  80 $\pm$  11 & 119 $\pm$  14  &  72 $\pm$  13  & 119 $\pm$  14  & 184 $\pm$  14 & 282 $\pm$  18  & 200 $\pm$  15
\\
M2  &  90 $\pm$  12  &  51 $\pm$   9  &  41 $\pm$   9 &  54 $\pm$  12  &  22 $\pm$  10  &  66 $\pm$  12  & 111 $\pm$  11 & 198 $\pm$  15  & 132 $\pm$  12
\\
M3  &  41 $\pm$  10  & $<$ 50  & $<$ 59  & $<$ 78  & $<$ 78 & $<$ 76  &  49 $\pm$   9  & 111 $\pm$  12  &  62 $\pm$   9
\\
M4  &  44 $\pm$  10  & $<$ 50  & $<$ 59  & $<$ 78  & $<$ 78 & $<$ 76  &  47 $\pm$   8  & 115 $\pm$  13  &  53 $\pm$   9
\\
M5  & $<$ 61  & $<$ 50  & $<$ 59  & $<$ 78  & $<$ 78 & $<$ 76  & $<$ 44  &  54 $\pm$  10  &  31 $\pm$   7
\\
M6  & $<$ 61  & $<$ 50  & $<$ 59  & $<$ 78  & $<$ 78 & $<$ 76  &  40 $\pm$   8  &  63 $\pm$  10  &  34 $\pm$   8
\\
M7  &  70 $\pm$  11  &  83 $\pm$  11  &  42 $\pm$   9  & $<$ 78  & $<$ 78  & $<$ 76  &  90 $\pm$  11  & 141 $\pm$  14  & 110 $\pm$  11
\\
M8  & $<$ 61  & $<$ 50  & $<$ 59  & $<$ 78  & $<$ 78  & $<$ 76  &  28 $\pm$   7  &  60 $\pm$  10  &  38 $\pm$   8
\\
M9  &  41 $\pm$  10  & $<$ 50  & $<$ 59  & $<$ 78  & $<$ 78  & $<$ 76  &  41 $\pm$   8  &  78 $\pm$  11  &  48 $\pm$   8
\\
\hline\\[-7pt]

$\Sigma$M$_{\rm i}$  & $402\pm66$ & $201\pm30$ & $163\pm29$ & $173\pm26$ & $94\pm23$ & $185\pm26$ & $590\pm76$ & $1102\pm113$ & $708\pm87$\\

Ring  &  $2023\pm81$ & $1336\pm65$ & $583\pm59$ & $1114\pm79$ & $^{(a)}$ & $1262\pm80$ & $2326\pm76$ & $3700\pm94$ & $2435\pm76$\\
  
\hline\\[-7pt]

   $\sigma$      & 37         &   30       &    27      &       47   &  47       &       46   &        27  &   35       &   27\\[5pt]

\hline

\end{tabular}
\end{center}
$^{(a)}$ It was not possible to perform aperture photometry because of insufficient signal-to-noise.
\end{table*}

The data were acquired in service mode queued observations under
stable weather conditions with an optical seeing of $\leq 0$\farcs8
and an air mass $\leq 1.5$. We selected the pixel scale of 0\farcs127
pixel$^{-1}$, which provides a total FOV of
32\farcs3$\times$32\farcs3. The observations were performed using the
standard chopping and nodding technique, which removes the
time-variable sky background, telescope thermal emission and most of
the so-called 1/$f$ noise.  We used a chopper throw of 30\arcsec. All
observations were bracketed by photometric standard star observations.
We note that the chopping/nodding technique, while effectively
reducing the background, may also eliminate any large-scale diffuse
nebular emission from the region. VISIR imaging, however, recovers
nearly 50\% of the total \NeII\ line flux measured by
\spitzer\ (Table~\ref{table:big}) and hence we do not consider
chopping/nodding to introduce a significant source of systematic
photometric errors.

The final images are the result of shifting and combining the
individual chopping cycles and were obtained using the VISIR pipeline
(version 1.3.7). Practically all of the final images showed some
stripes due to high-gain pixels.  We removed these stripes using the
reduction routine {\it destripe} (Snijders, private communication),
which is written in IDL.  The conversion from counts to physical flux
units was derived from the photometric standards HD\,216032 and
HD\,218670.  The angular resolution varies between 0\farcs3 and
0\farcs4, very close to the diffraction limit of VLT at 10\m.

  \begin{figure}
   \begin{minipage}[c]{0.333\columnwidth} 
      \centering \includegraphics[width=\textwidth]{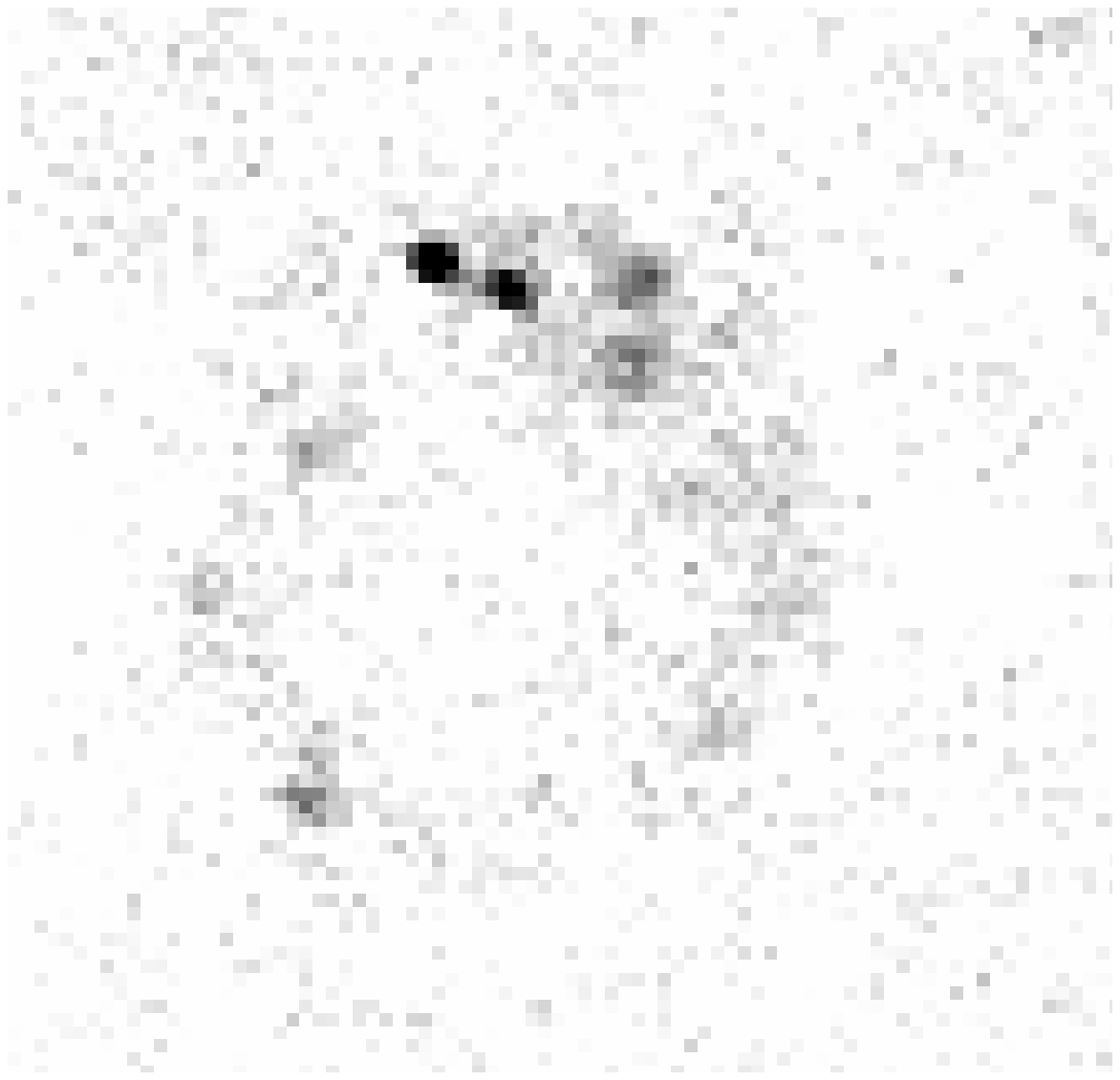}
   \end{minipage}%
   \begin{minipage}[c]{0.333\columnwidth} 
      \centering \includegraphics[width=\textwidth]{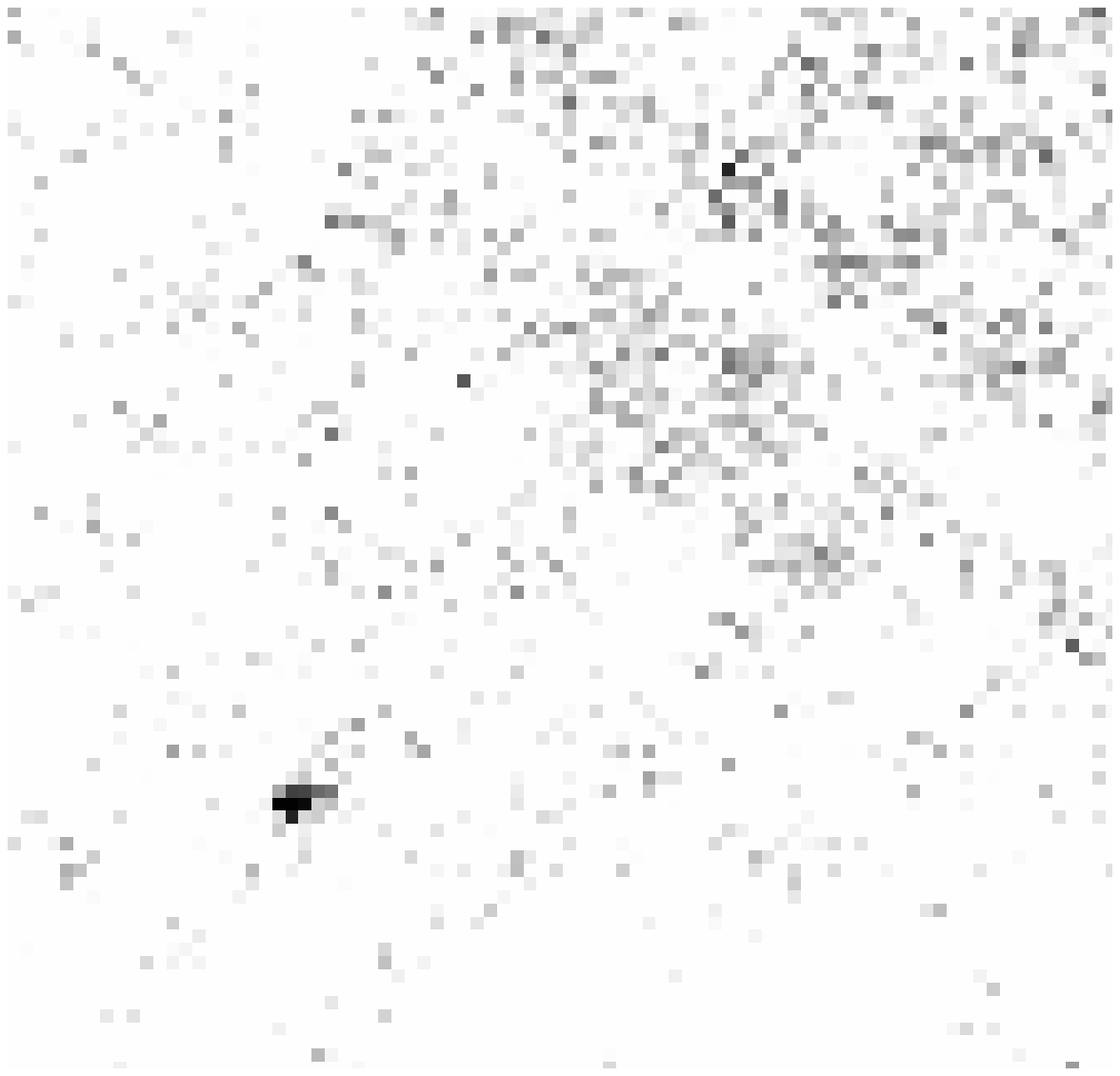}
   \end{minipage}%
   \begin{minipage}[c]{0.333\columnwidth} 
      \centering \includegraphics[width=\textwidth]{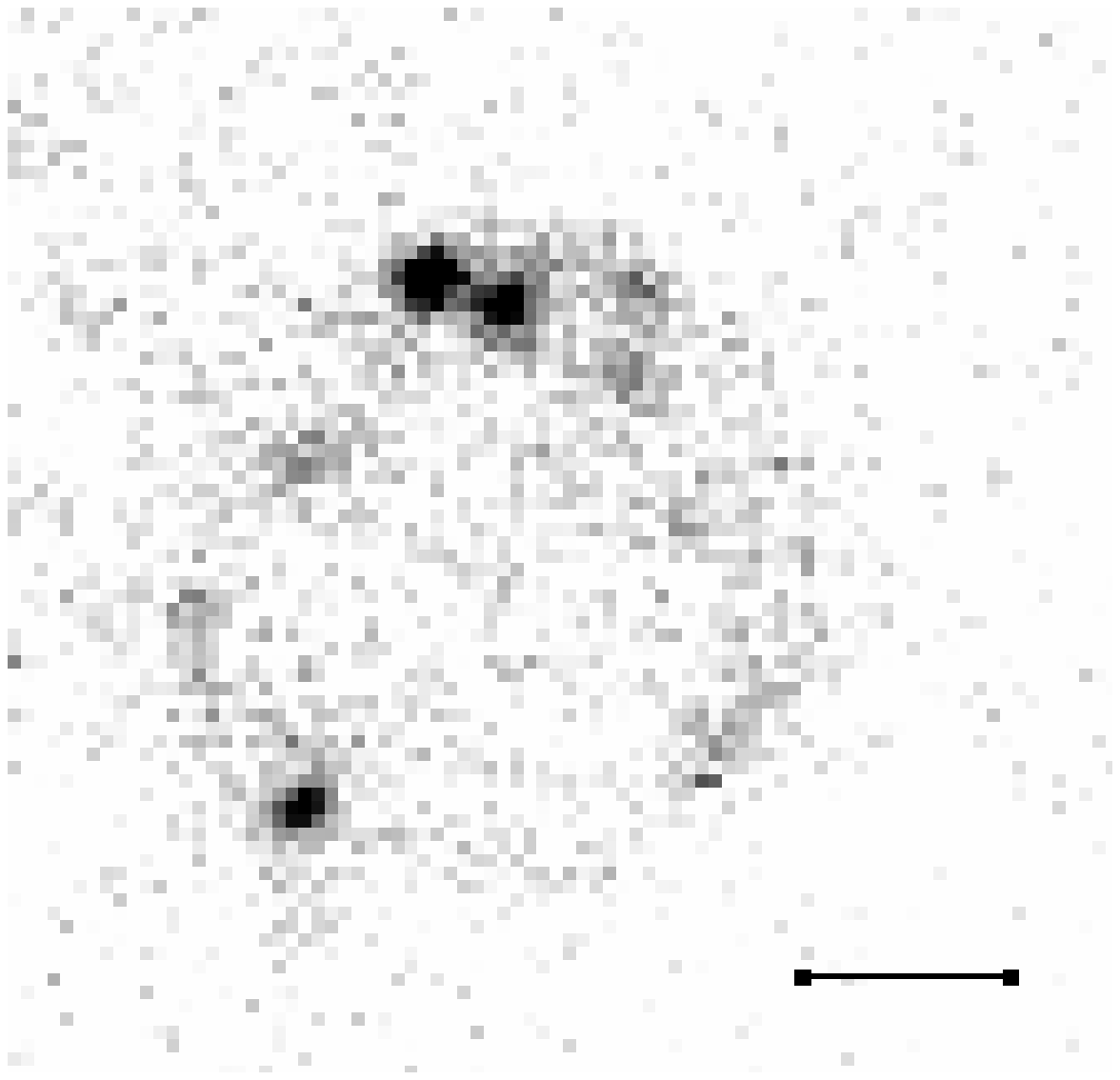}
   \end{minipage}%
     \caption{{\em Left:} Continuum subtracted \NeII[12.8\m] line
       image. {\em Center:} 8.6\m\ PAH image {\em Right:}
       12\m\ continuum images taken in the SiC filter.  The PAH image
       suffers from fixed pattern noise, but shows one bright peak,
       M7. All three images are displayed in linear intensity scaling;
       North is up and East is to the left.  The bar indicates a scale
       of $2''$ or 190\,pc.}
     \label{fig:raw}
  \end{figure}

Fig.~\ref{fig:raw} illustrates the image quality of the resulting maps
in the continuum subtracted \NeII[12.8\m] line, the 8.6\m\ feature of
polycyclic aromatic hydrocarbons (PAHs) and the MIR continuum.
Although the PAH map suffers from fixed pattern noise, the \NeII\ and
continuum maps illustrate the significant improvement in comparison to
previous mid-IR observations \citep[e.g., ][Fig. 4]{schinnerer97}.

\begin{table}
\caption{Absolute and relative coordinates of the MIR peaks with respect to the radio nucleus (Nuc) from \citet{forbes94}.}
\label{tbl:coord}
\begin{center}
\begin{tabular}{l@{\hspace{7pt}}c@{\hspace{7pt}}c@{\hspace{7pt}}c@{\hspace{7pt}}c@{\hspace{7pt}}c}
\hline\hline\\[-5pt]       
Source & $\alpha$ (J2000) & $\delta$ (J2000) & $\Delta \alpha (\arcsec)$ & $\Delta \delta(\arcsec)$ \\ 
\hline\\[-5pt]

Nuc & $23^{h} 16^{m} 10.\!^{s}71$ & $-42\deg 35\arcmin 04.\!\arcsec 4$ & $+0.0$ & $+0.0$ \\
M1        & $23^{h} 16^{m} 10.\!^{s}74$ & $-42\deg 35\arcmin 01.\!\arcsec 6$ & $+0.3$ & $+2.8$ \\
M2        & $23^{h} 16^{m} 10.\!^{s}67$ & $-42\deg 35\arcmin 01.\!\arcsec 8$ & $-0.4$ & $+2.6$ \\
M3        & $23^{h} 16^{m} 10.\!^{s}57$ & $-42\deg 35\arcmin 02.\!\arcsec 6$ & $-1.6$ & $+1.8$ \\
M4        & $23^{h} 16^{m} 10.\!^{s}56$ & $-42\deg 35\arcmin 01.\!\arcsec 8$ & $-1.7$ & $+2.6$ \\
M5        & $23^{h} 16^{m} 10.\!^{s}48$ & $-42\deg 35\arcmin 02.\!\arcsec 3$ & $-2.6$ & $+2.1$ \\
M6        & $23^{h} 16^{m} 10.\!^{s}49$ & $-42\deg 35\arcmin 06.\!\arcsec 1$ & $-2.5$ & $-1.7$ \\
M7        & $23^{h} 16^{m} 10.\!^{s}85$ & $-42\deg 35\arcmin 06.\!\arcsec 7$ & $+1.6$ & $-2.3$ \\
M8        & $23^{h} 16^{m} 10.\!^{s}94$ & $-42\deg 35\arcmin 04.\!\arcsec 8$ & $+2.5$ & $-0.4$ \\
M9        & $23^{h} 16^{m} 10.\!^{s}84$ & $-42\deg 35\arcmin 03.\!\arcsec 4$ & $+1.5$ & $+1.0$ \\[5pt]

\hline
\end{tabular}
\end{center}

\end{table}

\begin{figure*}
    \centering
    \begin{minipage}[c]{0.5\textwidth} 
      \centering \includegraphics[width=\textwidth]{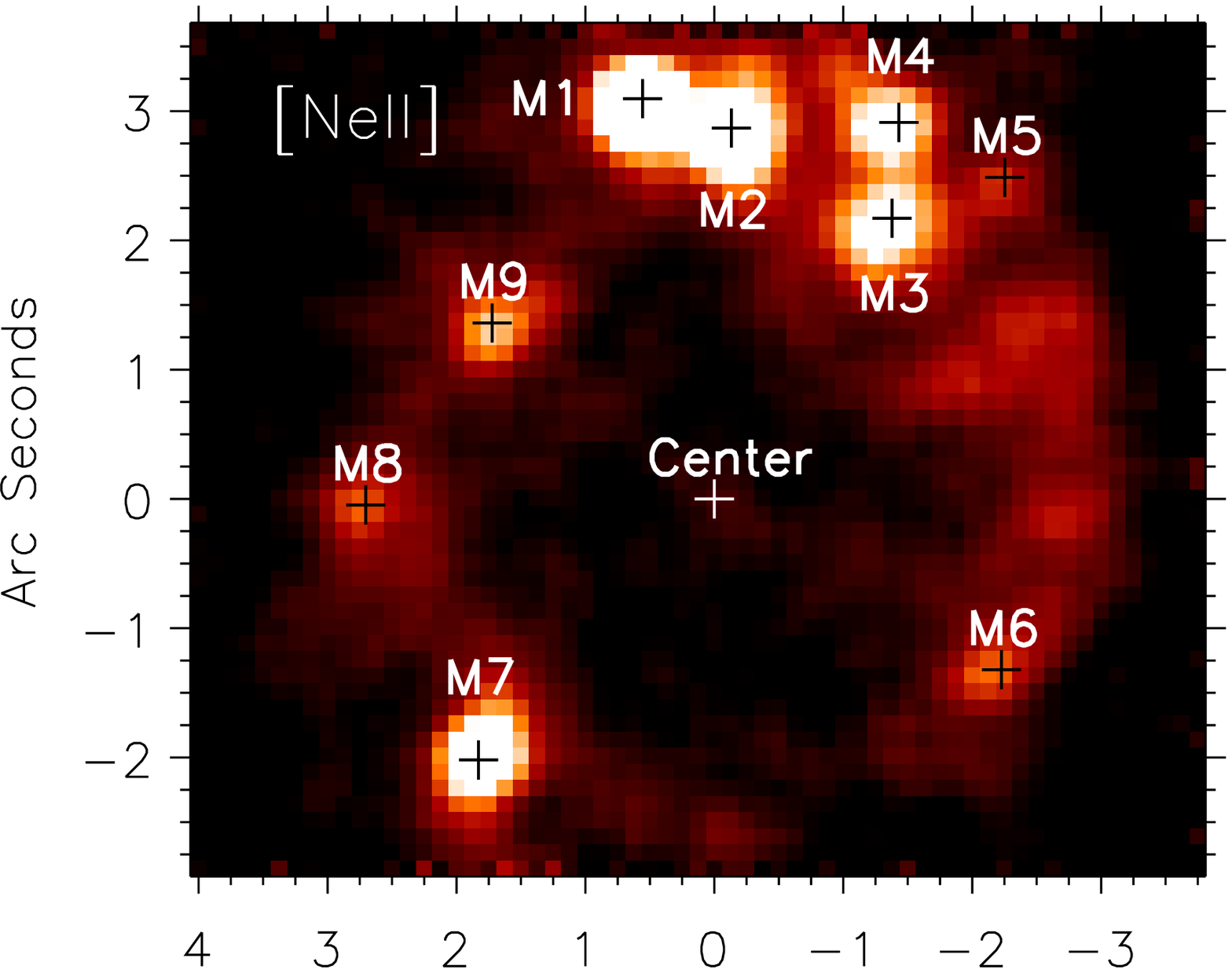}
    \end{minipage}%
   \begin{minipage}[c]{0.5\textwidth} 
      \centering \includegraphics[width=\textwidth]{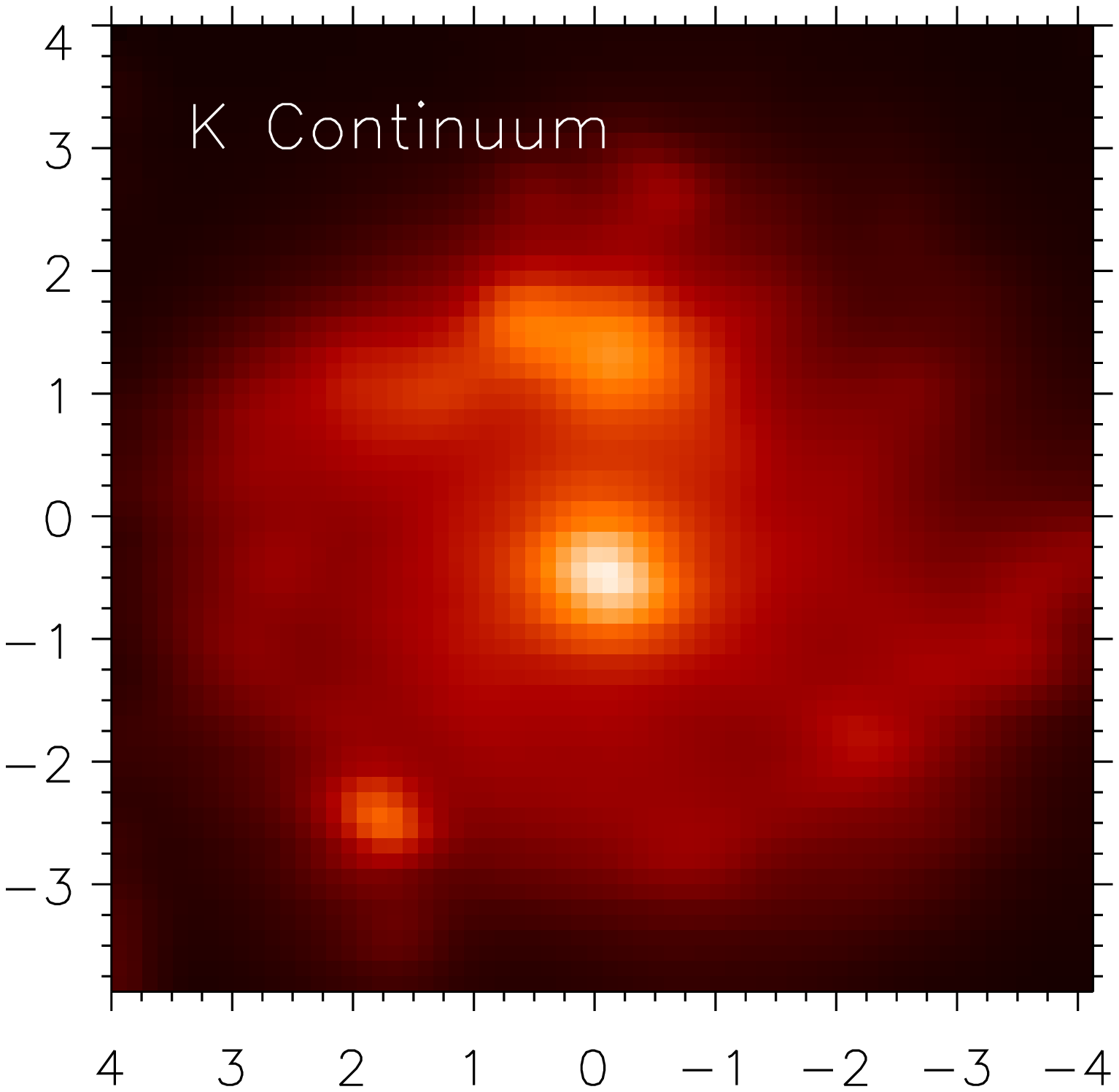}
    \end{minipage} \\
   \begin{minipage}[c]{0.5\textwidth} 
      \centering \includegraphics[width=\textwidth]{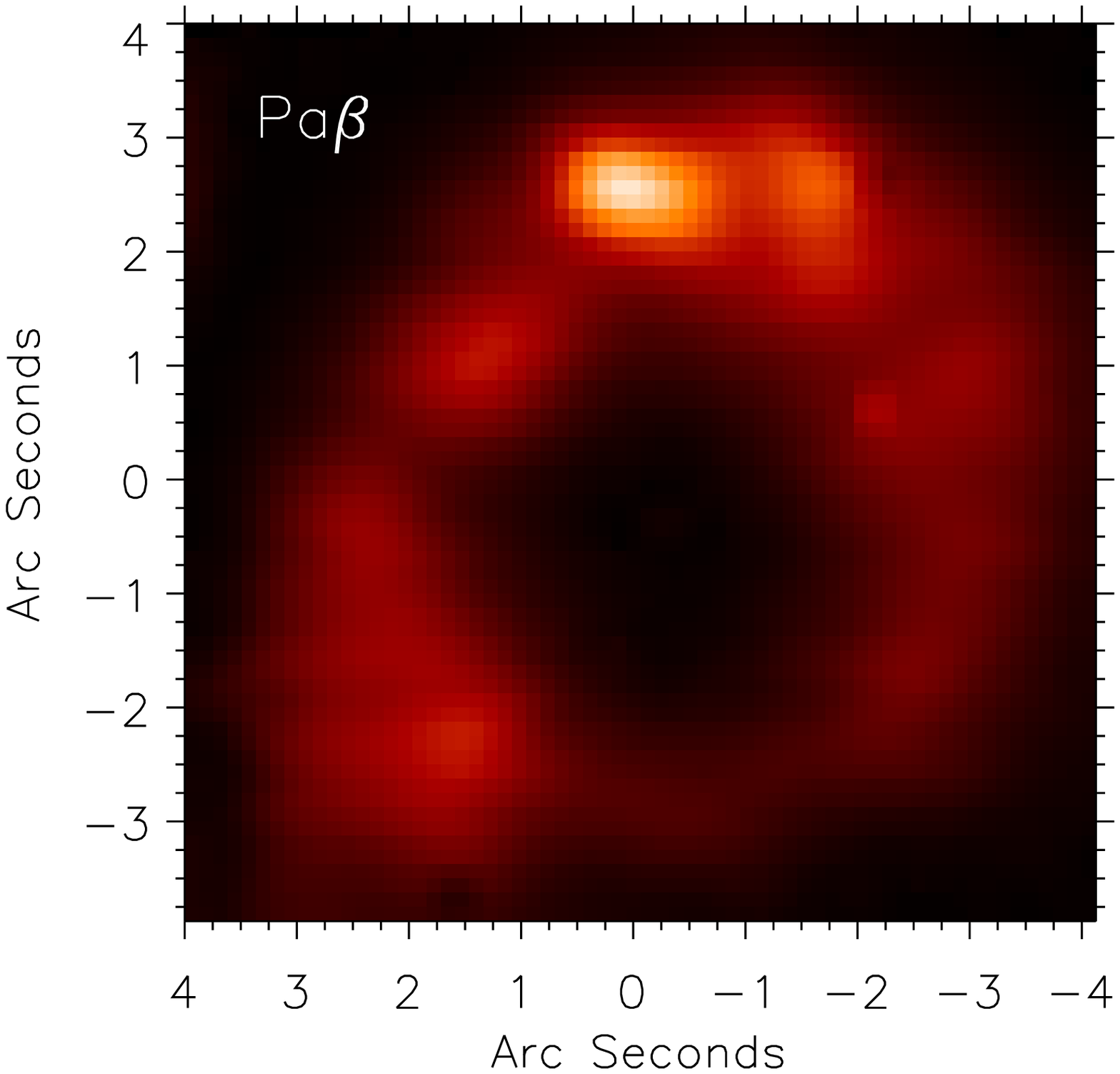}
    \end{minipage}%
   \begin{minipage}[c]{0.5\textwidth} 
      \centering \includegraphics[width=\textwidth]{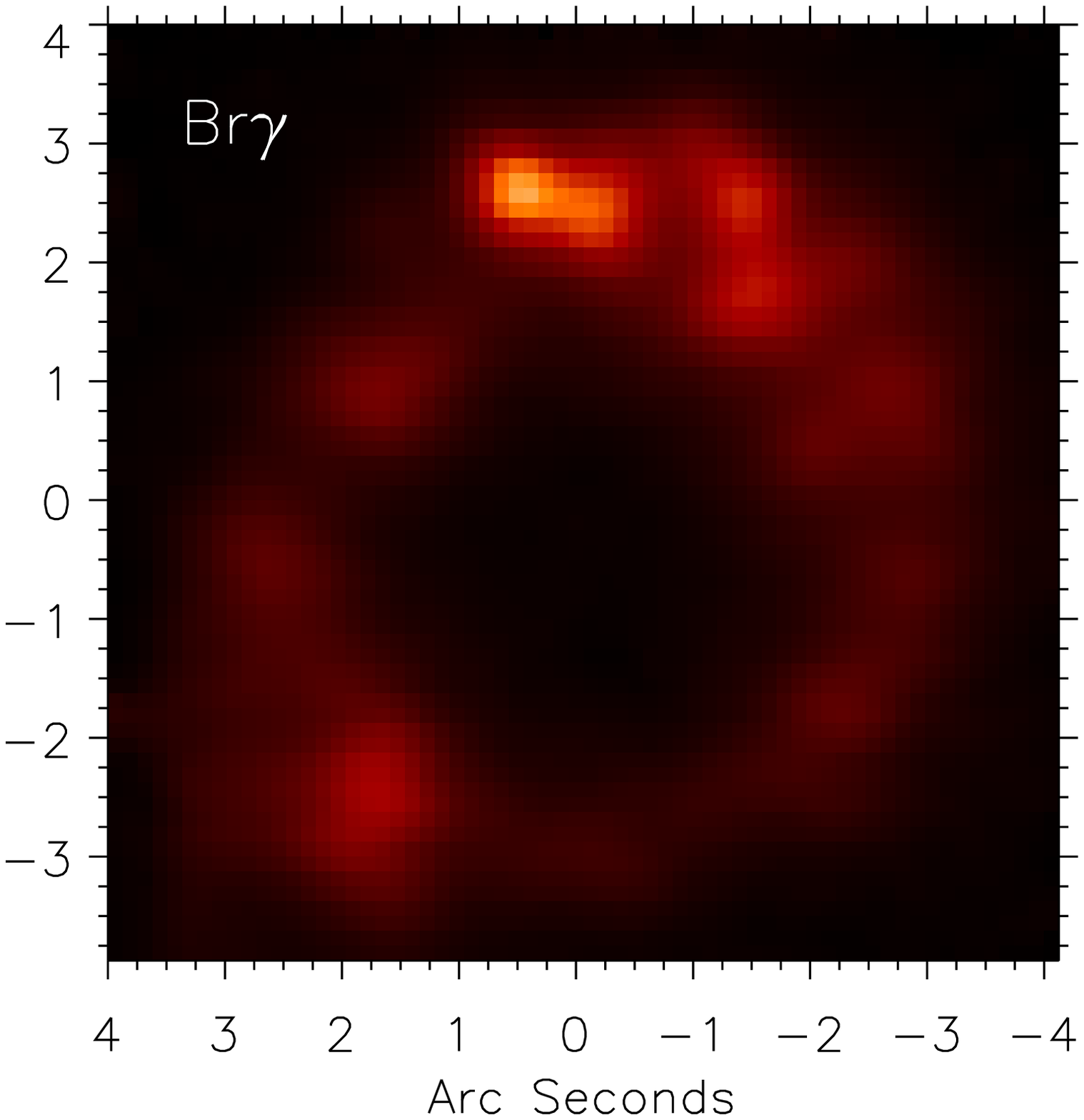}
    \end{minipage}
    \caption{ {\em a) upper left:} VISIR image taken in the
      12.8\m\ filter. The $1-\sigma$ RMS is approximately 0.57\,mJy
      pixel$^{-1}$.  The plus signs indicate the locations of the MIR
      peaks M1 through M9, discussed in this paper.  {\em b) upper
        right:} K-band continuum image of NGC~7552, composed from the
      reconstructed SINFONI data cube.  {\em c) lower left:}
      Reconstructed and continuum subtracted SINFONI \pab\ image.
      {\em d) lower right:} Reconstructed and continuum subtracted
      SINFONI \brg\ image.  All four figures indicate the image scale
      in arcseconds ($1\arcsec$ corresponds to 95\,pc) with respect to
      the radio continuum center (see Table~\ref{tbl:coord} and
      section~\ref{relastrometry}).}
     \label{fig:fourplots}
  \end{figure*}

\begin{figure*}
   \begin{minipage}[c]{0.5\textwidth} 
      \centering \includegraphics[width=\textwidth]{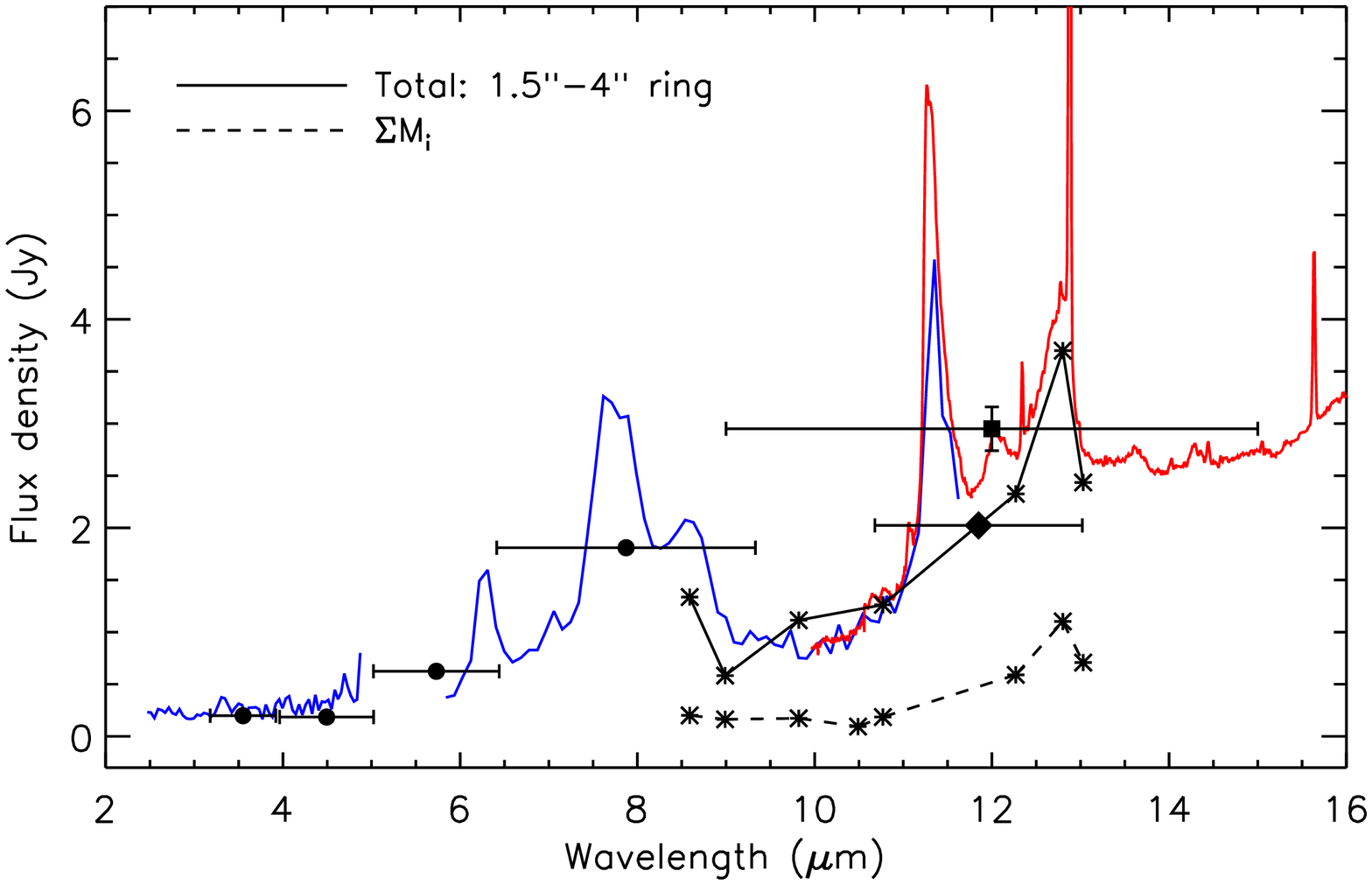}
   \end{minipage}
   \begin{minipage}[c]{0.5\textwidth}
      \centering \includegraphics[width=\textwidth]{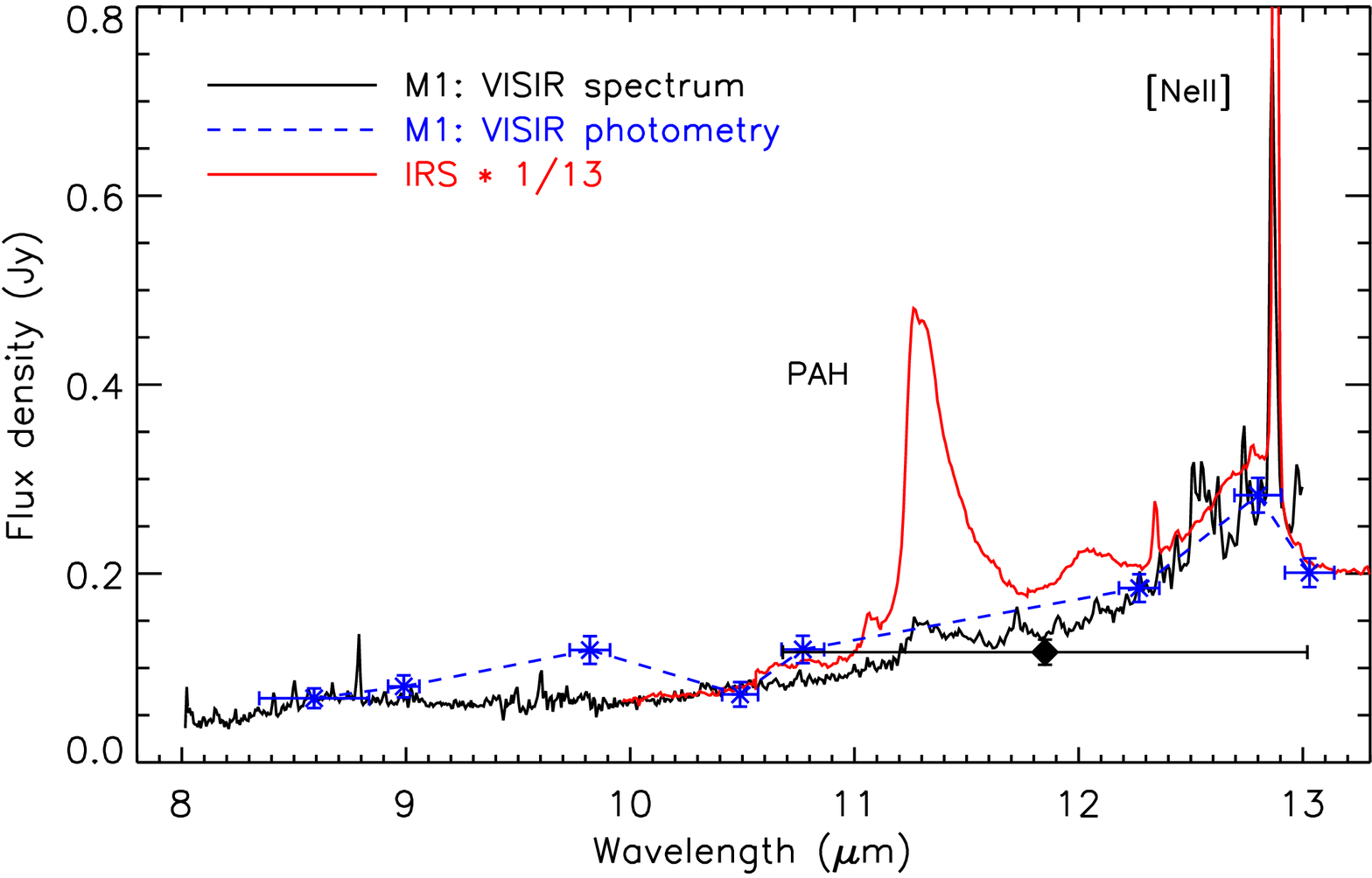}
   \end{minipage}
   \caption{{\em Left:} The mid-IR SED of the total integrated flux
     within the $1.\!''5\leq r \leq 4''$ ring (black solid line) and
     the sum of our nine peaks $\Sigma$M$_i$ (black dashed line). The
     integrated flux of the entire galaxy as measured through the SiC
     filter (Table~\ref{table:fluxes}) is indicated by the diamond.
     The filled circles show the IRAC photometry using a 12\farcs2
     aperture. The filled square is the 12\m\ IRAS flux taken from
     \citet{gil06}.  The spectra from ISOPHOT (blue line) and
     \spitzer-IRS (red line) were taken through apertures of size
     $24''\times 24''$ and $27''\times 43''$, respectively.
     {\em Right:} The VISIR low-resolution N~band spectrum of peak
     M1 (black line) in comparison to the \spitzer-IRS spectrum (red
     line).  The latter covers the central $27''\times 43''$ of
     NGC\,7552 and was normalised to the former at 10\m\ (scaled down
     by a factor of 13).  In both spectra, the \NeII\ fine-structure
     line at 12.8\m\ is clearly present but the strength of the PAH
     11.3\m\ band differs significantly.  In addition, the blue
     crosses indicate the VISIR narrowband photometry of M1 and the
     black diamond corresponds to the broadband SiC filter
     measurement.}
   \label{fig:sed}
  \end{figure*}

The starburst ring and several individual knots are clearly visible in
the 12.8\m\ filter image in Fig.~\ref{fig:fourplots}\,a.  We have
identified nine unresolved peaks, which we label M1 to M9. The
positions of these peaks are listed in Table~\ref{tbl:coord}.  For
each VISIR filter, the fluxes of each knot were measured performing
aperture photometry within circular apertures with a diameter of
0\farcs7, about twice the resolution limit. The photometric fluxes are
all tabulated in Table~\ref{table:fluxes}. M1 and M2 are the brightest
knots, clearly detected in all the VISIR filters.  The quoted
uncertainties are those derived from the aperture photometry. The
summed contribution of all the individual knots ($\Sigma$M$_{\rm i}$)
and the total photometry obtained within an
3\arcsec--8\arcsec\ annulus that covers the starburst ring
(section~\ref{starburstring}) are also listed.

The \NeII\ 12.8\m\ line emission of the individual clusters has been
estimated from the `NeII\_1', `NeII' and `NeII\_2' narrow-band
filters.  We interpolated the continuum level at the central
wavelength of the `NeII' filter (12.8\m) from the `NeII\_1' and
`NeII\_2' filters. The resulting \NeII\ line fluxes are listed in
Table~\ref{table:big}.

Similarly, we have obtained approximate measurements of the
8.6\m\ emission feature of PAHs.  The three most luminous peaks in
the `PAH\_1' filter are the mid-IR sources M1, M2 and M7.  Here we
have only used the `ArIII' filter to measure the continuum level and
assumed a flat continuum in $F_\nu$.  The results are listed in
Table~\ref{table:pahs}.  With only one data point to constrain the
baseline image the fluxes are more uncertain and should be compared to
spectroscopic measurements, where available.


\subsubsection{Spectroscopy}
\label{secspectroscopy}

A low-resolution spectrum at N-band was obtained of the brightest
peak M1 on June 16th, 2005.  The slit with a width of 0\farcs75 was
centred on M1 and oriented along P.A.=$-$10\degr.  The spectral
resolution is $R\thickapprox 185-390$.  Four spectral settings
($\lambda_{\rm centre}$ = 8.8, 9.8, 11.4, and 12.4\m), overlapping by
at least by 15\%, were used to cover the full N-band.  The total
on-source integration time was 30 minutes for each spectroscopic
setting. Since the images show no large-scale diffuse emission,
chopping and nodding on-slit were applied with a 12\arcsec\ chopper
throw. The early-type A0 stars HIP\,109268 and HIP\,113963 were
observed before and after each target observation for
spectro-photometric flux calibration.

We reduced the data with the VISIR pipeline, which includes
subtracting the chopped/nodded pairs, correcting for optical
distortion, wavelength calibration and extracting the spectra from a
1\farcs27 aperture in the spatial direction.  Absolute flux
calibration was obtained by integrating the standard star spectra over
the VISIR filter bands and normalising their fluxes to the ones of the
photometric standards.  The accuracy of the flux calibration is
approximately 20\%.  The resulting spectrum is shown in
Fig.~\ref{fig:sed}.


\subsection{SINFONI}

\subsubsection{Data Analysis}

Observations of NGC 7552 were also taken with the Spectrograph for
INtegral Field Observations in the Near Infrared (SINFONI) at the VLT.
SINFONI provides spatial and spectral data in the form of a data cube
covering the J, H, and K~bands.  The SINFONI instrument is mounted at
the Cassegrain focus of the VLT Unit Telescope 4, Yepun
\cite[e.g.][]{bonnet04,gillessen05}.  The observations were made in
the J, H, and K~bands using a spatial resolution setting of 0\farcs25
pixel$^{-1}$, corresponding to a field of view of
8\arcsec$\times$8\arcsec, and were obtained on August 27th, 2005
without adaptive optics correction.  The average air mass at the time
of observation was 1.23 and the DIMM seeing between $1''$ and
1\farcs3.  Each galaxy had an integration time of 300 seconds per band
and a spectral resolution $R$ of 2000, 3000, and 4000, for the J, H,
and K~bands respectively.

The reduction of the data was performed with the SINFONI pipeline
\citep{modigliani07}, version 1.7.1. This included the removal of
detector signatures (geometric distortions, bad pixels, pixel gain
variations, etc.), the sky emission correction, the wavelength
calibration and the image reconstruction from the image slices. Flux
calibration and removal of the contamination from telluric lines was
performed using observations of the standard star HIP\,025007.  In
Figure~\ref{fig:fourplots}\,c,d we show the continuum subtracted
\pab\ and \brg\ line maps and representative sections of the J and K
spectra to indicate the signal-to-noise.  Although the H band was
included in our observations, it is suffering heavily from OH
atmospheric line contamination and not relevant for the following
analysis.

We have also constucted the K~band continuum image from integrating
the SINFONI spectra across the K~band without including the emission
from strong lines. The result is shown in Fig.~\ref{fig:fourplots}\,b
and discussed in section~\ref{starburstring}.

Table~\ref{table:big} lists the derived Pa$\beta$ and \brg\ line
fluxes, and the equivalent widths (EWs) of \brg, averaged over
circular apertures of 0\farcs7 centred on each mid-IR peak.  The
integrated line fluxes within the 3\arcsec--8\arcsec\ annulus are also
given in Table~\ref{table:big}. We emphasize that our measured
\brg\ line flux of $6.55\times 10^{-21}$~W\,cm$^{-2}$, integrated over
the central region, is in excellent agreement with the corresponding
values from previous studies by different authors:
$6.5\times 10^{-21}$~W\,cm$^{-2}$ \citep{moorwood90}, 
$6.4\times 10^{-21}$~W\,cm$^{-2}$ \citep{forbes94b}, and 
$5.7\times 10^{-21}$~W\,cm$^{-2}$ \citep{schinnerer97}.

  \begin{figure}
    \centering
    \resizebox{\hsize}{!}{\includegraphics{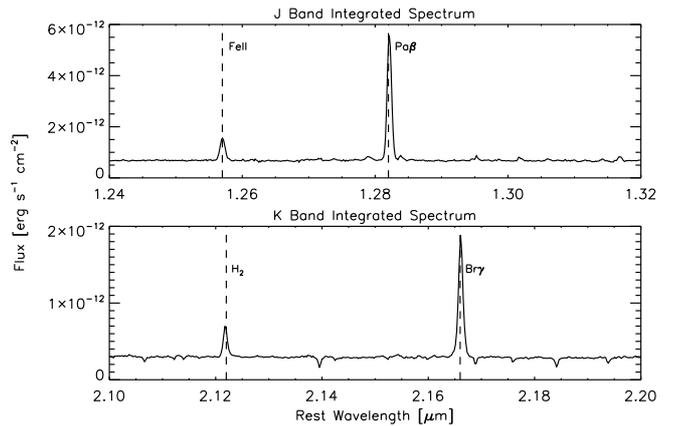}}
     \caption{Segments of the spatially integrated spectra at J ({\em
         top}) and K~band ({\em bottom}) to illustrate the
       signal-to-noise. The strongest emission lines (\FeII, \pab,
       H$_2$ and \brg\ are labelled.}
     \label{fig:SINFONI_JK}
  \end{figure}

\subsubsection{Uncertainties}

We estimate the uncertainty in the measured \pab\ and \brg\ line
fluxes to be 15\% (Table~\ref{table:big}).  This uncertainty is a
combination of the typically 10\% absolute uncertainty on the SINFONI
line fluxes -- based on the residual slope in the standard star
spectra after the division by the appropriate black body -- and the
systematic uncertainty in the peak position (estimated from shifting
the extraction apertures by $\pm 1$ spaxel).

The uncertainty of the EW of \brg\ is only 5\% for two reasons.
First, because an EW is a ratio of two measurements at similar
wavelenths, the absolute photometric errors cancel out.  Second,
because the EW changes (with respect to the line flux) less rapidly
with distance from the position of the IR peaks the EW is less
sensitive to the positioning of the extraction aperture.  The EW of
\brg\ is mainly of importance for the derivation of the cluster ages .
We note that all of the measurement uncertainties are small in
comparison to the systematic errors given by the model assumptions
(see sections \ref{sec:hydrogenlines} and \ref{sect:ages}).

\setcounter{table}{2}
\begin{table*}
\caption{The physical properties of the clusters M1 - M9 derived from 0\farcs7 apertures centered on the mid-IR peaks: Pa$\beta$, Br$\gamma$ (both measured and extinction corrected), equivalent widths of Br$\gamma$, and the \NeII\ line, and 12\,\m\ continuum (NeII\_1 filter) fluxes.  The derived values of extinction $A_V$, number of Lyman-continuum photons $N_{Lyc}$, the equivalent number of O7V stars, the total infrared luminosity, and the approximate cluster ages are also listed.  For a discussion of the quoted uncertainties see section~\ref{obs}.}
\label{table:big}
\begin{center}
\begin{tabular}{l@{\hspace{0pt}}ccc@{\hspace{5pt}}c@{\hspace{5pt}}c@{\hspace{3pt}}c@{\hspace{3pt}}ccc@{\hspace{5pt}}c@{\hspace{5pt}}c}
\hline\hline\\[-7pt]
Source & Pa$\beta$ & Br$\gamma$ & (Br$\gamma$)$_{\rm corr}$ & EW(Br$\gamma$) & \NeII\ & 12\,\m\ cont. & $A_V$ & $N_{Lyc}$ & $N_{\rm O7V}$ & $L_{\rm IR}$ & Age \\
& \multicolumn{3}{c}{($10^{-23}$\,W\,cm$^{-2}$)} & (\AA) & ($10^{-21}$\,W\,cm$^{-2}$) & (mJy) & (mag) & ($10^{51}$ s$^{-1}$) & ($\times10^3$) & ($10^9$\lsun) & (Myr)\\
\hline\\[-5pt]

M1 & $27.1\pm4.1$ & $12.4\pm1.9$ & $23.5\pm4.9$ & $65.5\pm3.3$ & $25\pm9$ & $184\pm14$ & $8.1$ & $7.8\pm1.6$ & $1.4\pm0.3$ & 6.0 & $5.6$\\
M2 & $22.1\pm3.3$ &  $9.6\pm1.4$ & $17.6\pm3.7$ & $45.8\pm2.3$ & $21\pm7$ & $111\pm11$ & $7.6$ & $5.8\pm1.2$ & $1.0\pm0.2$ & 4.0 & $5.8$\\
M3 & $15.1\pm2.3$ &  $7.7\pm1.2$ & $15.4\pm3.2$ & $42.0\pm2.1$ & $15\pm5$ & $49\pm9$   & $9.0$ & $5.1\pm1.1$ & $0.9\pm0.2$ & 1.9 & $5.9$\\
M4 & $22.4\pm3.4$ &  $9.8\pm1.5$ & $17.9\pm3.8$ & $74.5\pm3.7$ & $18\pm5$ & $47\pm8$   & $7.7$ & $5.9\pm1.2$ & $1.1\pm0.2$ & 1.6 & $5.5$\\
M5 & $10.3\pm1.5$ &  $4.4\pm0.7$ &  $7.9\pm1.7$ & $43.9\pm2.2$ &  $7\pm4$ & $<44$      & $7.5$ & $2.6\pm0.5$ & $0.5\pm0.1$ & 0.9 & $5.8$\\
M6 & $11.3\pm1.7$ &  $4.6\pm0.7$ &  $8.2\pm1.7$ & $15.4\pm0.8$ &  $8\pm5$ & $40\pm8$   & $7.1$ & $2.7\pm0.6$ & $0.5\pm0.1$ & 1.0 & $6.3$\\
M7 & $17.7\pm2.7$ &  $7.9\pm1.2$ & $14.7\pm3.1$ & $20.4\pm1.0$ & $11\pm6$ & $90\pm11$  & $7.9$ & $4.9\pm1.0$ & $0.9\pm0.2$ & 3.3 & $6.2$\\
M8 & $13.3\pm2.0$ &  $4.5\pm0.7$ &  $7.0\pm1.5$ & $15.5\pm0.8$ &  $7\pm5$ & $28\pm7$   & $5.6$ & $2.3\pm0.5$ & $0.4\pm0.1$ & 1.1 & $6.3$\\
M9 & $12.1\pm1.8$ &  $3.9\pm0.6$ &  $5.8\pm1.2$ & $16.8\pm0.8$ &  $9\pm5$ & $41\pm8$   & $5.2$ & $1.9\pm0.4$ & $0.3\pm0.1$ & 1.4 & $6.3$\\[2pt]
\hline\\[-7pt]
$\Sigma$M$_{\rm i}$& $151.4\pm22.7$ & $64.8\pm9.7$ & $118.0\pm24.8$ & & $121\pm17$ & $590\pm76$ & $7.4$ & $39.1\pm8.2$ & $7.0\pm1.5$ & 21.3 & $5.9\pm0.3$\\
Ring & $2020.8\pm303.1$ & $654.8\pm98.2$ & $1002.0\pm210.4$ & & $382\pm20$ & $2326\pm76$ & $5.3$ & & & \\[2pt]
\hline
\end{tabular}
\end{center}

\end{table*}


\subsection{Spitzer-IRAC}

\begin{figure*}
   \begin{minipage}[c]{0.25\textwidth} 
      \centering \includegraphics[width=\textwidth]{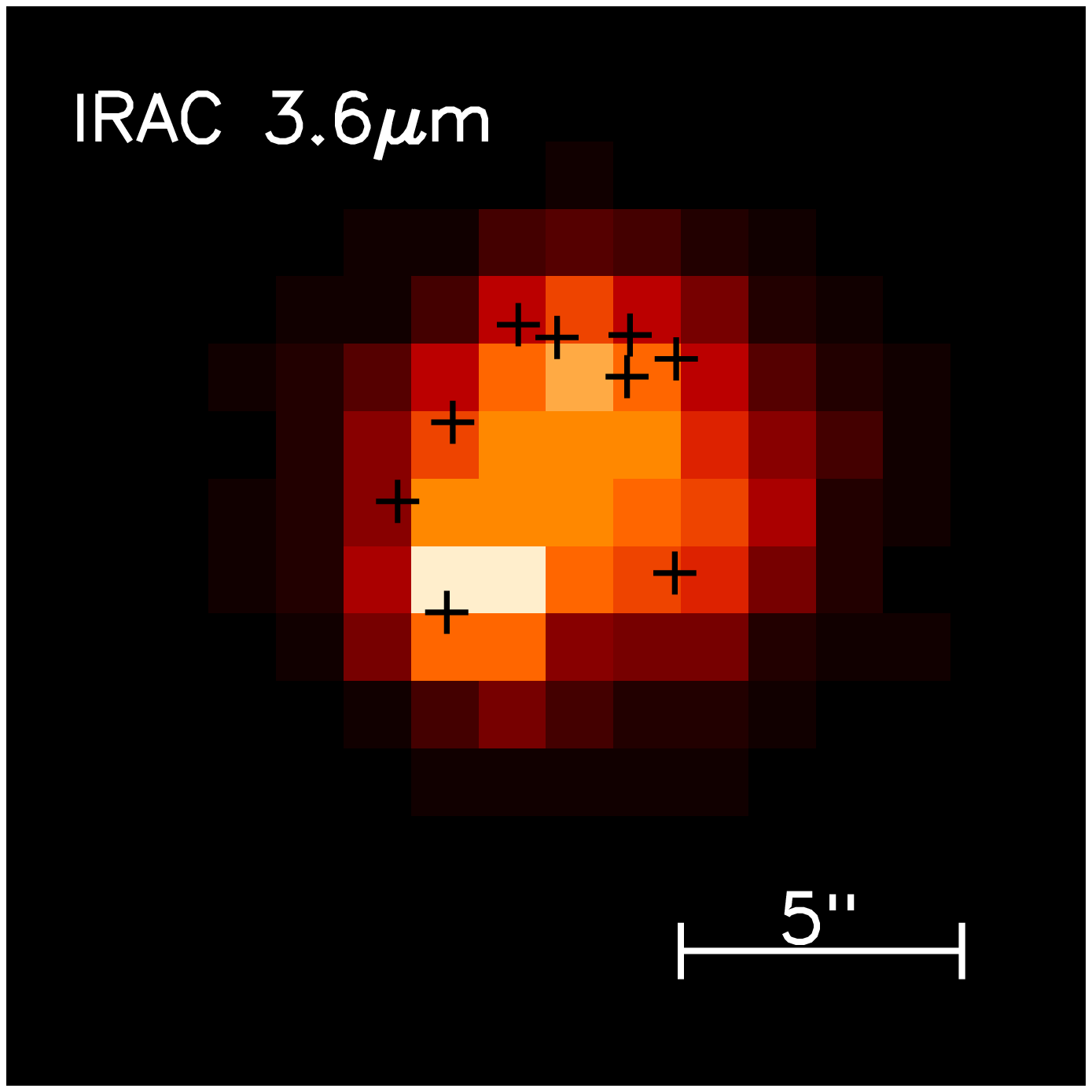}
   \end{minipage}%
   \begin{minipage}[c]{0.25\textwidth}
      \centering \includegraphics[width=\textwidth]{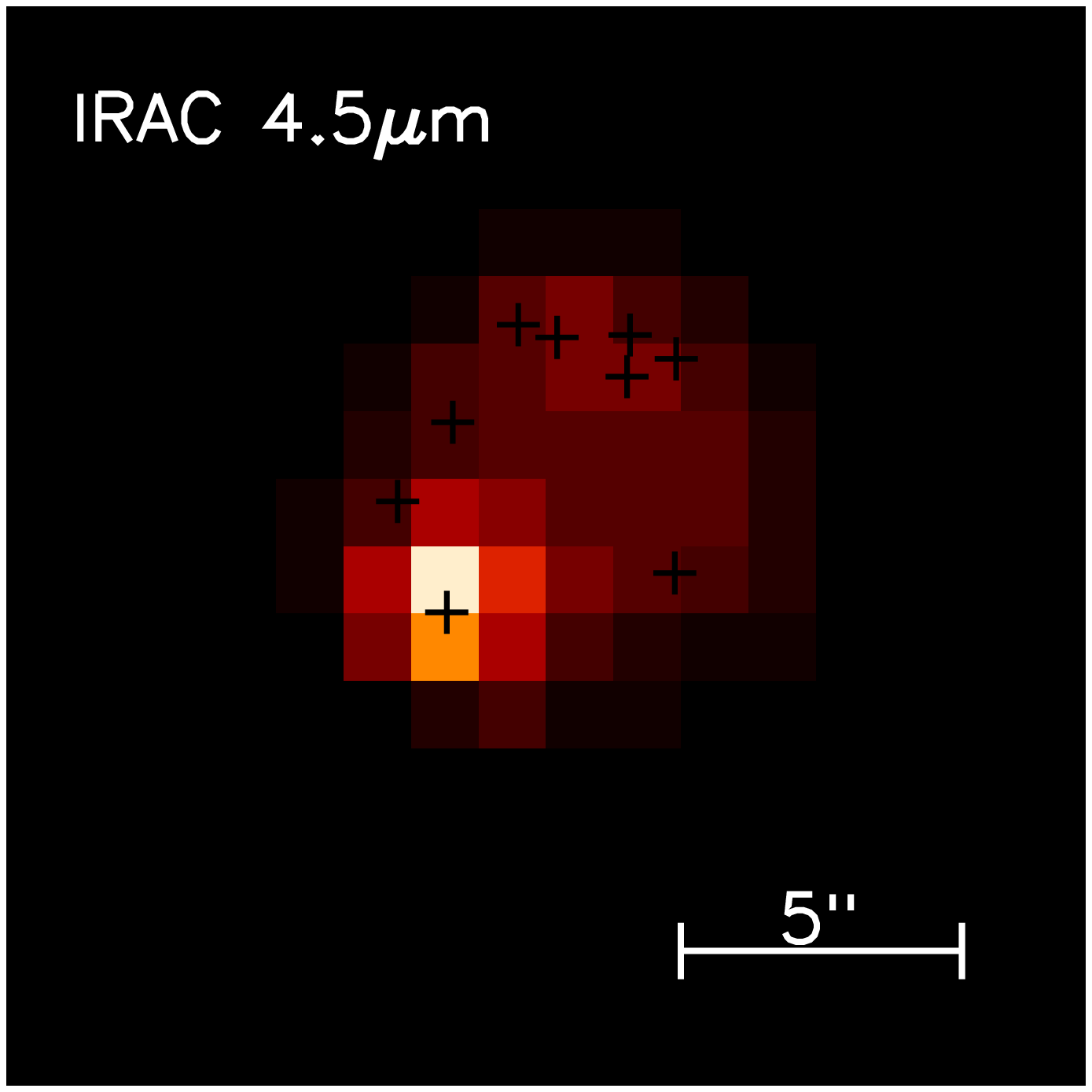}
   \end{minipage}%
    \begin{minipage}[c]{0.25\textwidth} 
      \centering \includegraphics[width=\textwidth]{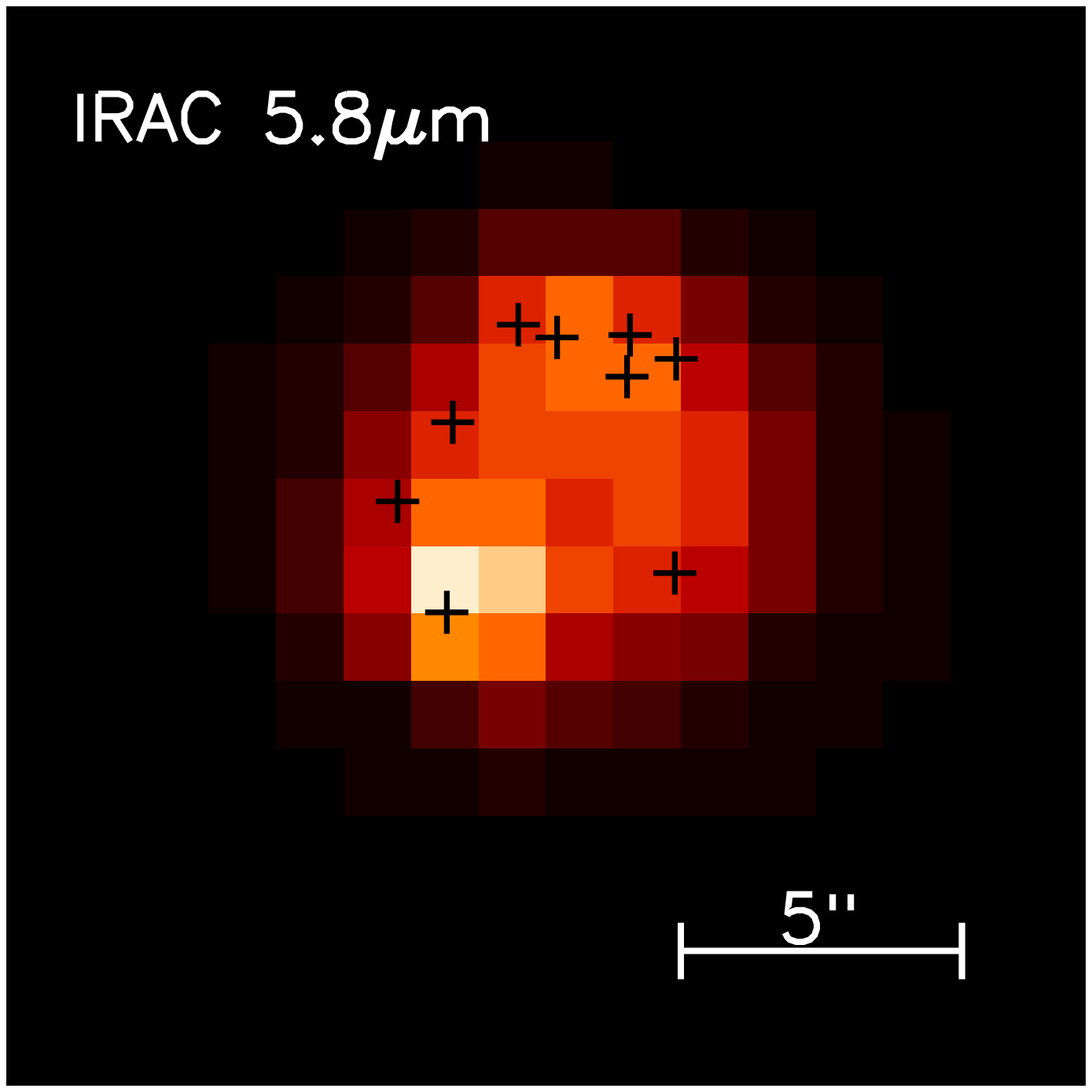}
   \end{minipage}%
   \begin{minipage}[c]{0.25\textwidth}
      \centering \includegraphics[width=\textwidth]{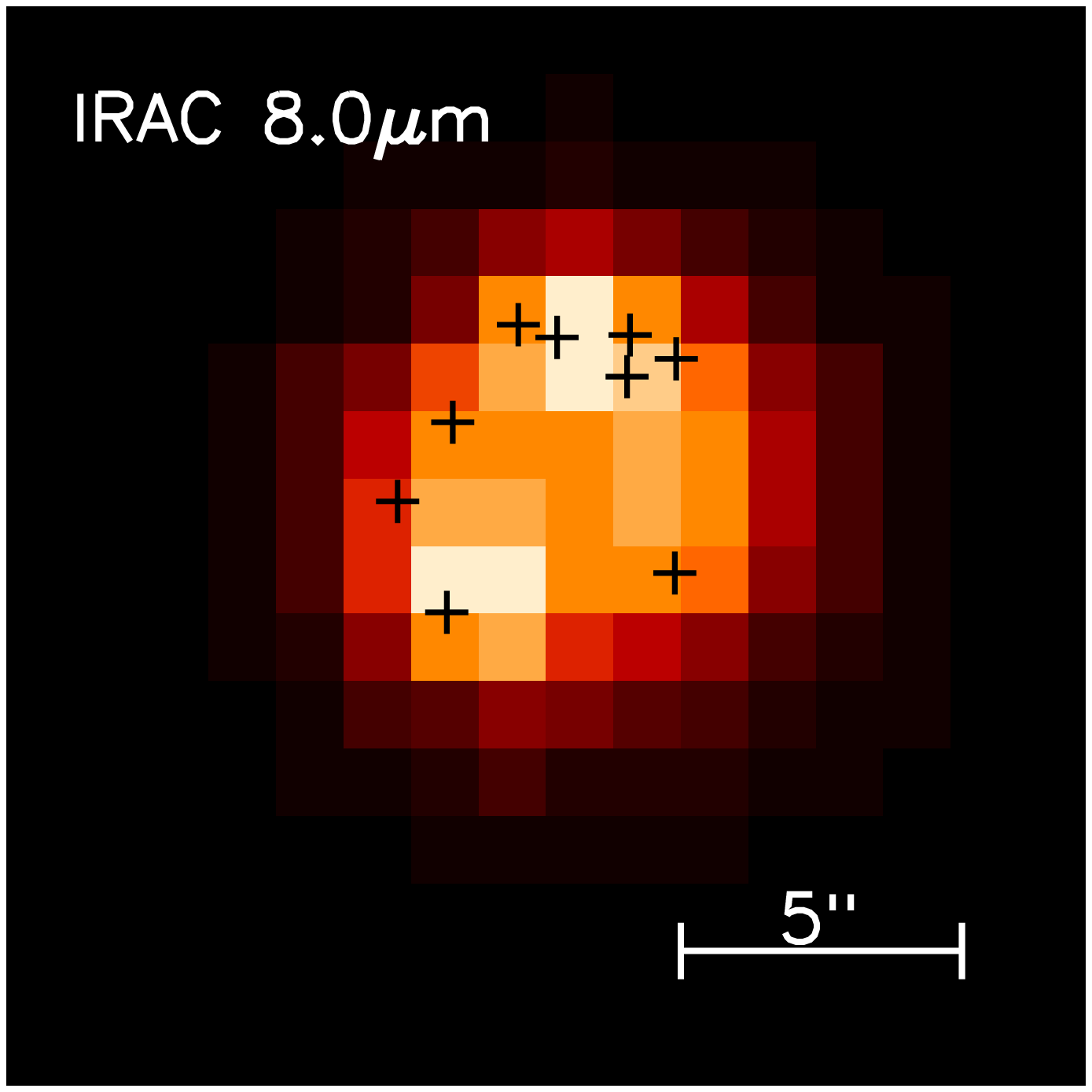}
   \end{minipage}
   	\caption{\spitzer-IRAC images of NGC\,7552 in the 3.6, 4.5,
          5.8 and 8.0\m\ bands. The plus signs indicate the positions
          of the VISIR MIR sources.}
   	\label{fig:irac}
  \end{figure*}

We have also analysed archival IRAC/\spitzer\ images of NGC\,7552 from
the SINGS Legacy Science Program \citep[e.g.][]{dale05}.  The images
are shown in Fig.~\ref{fig:irac}.  Due to the smaller
telescope aperture the angular resolution is much lower, approximately
$\sim$2\farcs4.  Nevertheless, two distinct, bright peaks in the ring
are clearly visible.  We have also performed aperture photometry using
the standard IRAC 10~pixel (12\farcs2) aperture radius with a sky
annulus of $10 - 20$ pixels.  According to the IRAC Data Handbook
(version 3.0) the photometric errors are dominated by the absolute
calibration uncertainty of 10\% resulting from the IRAC filter
bandpass responses in sub-array mode \citep{quijada04,reach05}. The
derived photometric fluxes are 198\,mJy, 185\,mJy, 624\,mJy and
1809\,mJy for IRAC bands 1 through 4, respectively.  These flux
densities are also indicated in Fig.~\ref{fig:sed}.

\subsection{Relative Astrometry}
\label{relastrometry}

Our analysis is based on multi-wavelength data, and the co-alignment
and relative astrometry of the various maps is of utmost importance,
as illustrated in Fig.~\ref{fig:whereclusters}.  Our starting
  point is the radio map of \citet{forbes94} with source B (radio
  nucleus, Table~\ref{tbl:coord}) as reference point.  The absolute
  astrometry of the VISIR images was then derived from shifting the
  \NeII\ image to best match the radio map.  Next, we compared the
  morphology of the \brg\ emission (Fig.~\ref{fig:fourplots}\,d) with
  the \NeII\ map (Fig.~\ref{fig:fourplots}\,a) and found an excellent
  agreement concerning the cluster peak positions.  We then shifted
  the SINFONI maps by a small amount (within the VLT pointing
  uncertainties) to match the \brg\ and \NeII\ maps.

The adjusted K~band continuum image (Fig.~\ref{fig:fourplots}\,b)
shows the K~band nucleus at RA $=23^{h} 16^{m} 10.\!^{s}69; \delta
=-42\deg 35\arcmin 04.\!\arcsec 8$, which is about $0.\!\arcsec 5$ to
the South from the radio nucleus.  We emphasize that the offset
between apparent K-band nucleus and radio center (see also
Fig.~\ref{fig:fourplots}) is bootstrapped from the \brg\ image, which
is part of the IFU data from which the K-band continuum was
reconstructed.

The HST images shown in Fig.~\ref{fig:whereclusters} were combined on
the basis of the HST reconstructed pointing, and the \NeII
12.8\m\ contours have been adjusted to the radio map (as described
above) and then overplotted onto the HST coordinate frame.  Like the
K-band continuum, the HST images show the nucleus slightly offset to
the South with respect to the active starburst ring.  However, even if
this offset would indicate the uncertainty in the absolute astrometric
registration, the scientifically more relevant uncertainty here is the
{\em relative} astrometry between emission peaks, which is
approximately two VISIR pixels or $0\farcs25$.


\section{Analysis}
\label{analysis}


\subsection{Appearance of the Ring at Different Wavelengths}
\label{starburstring}

As mentioned in the Introduction, the morphology of the inner region
of NGC\,7552 is quite complex.  Based on imaging at BVI bands,
\citet{feinstein90} found a nuclear region of low extinction within
$4\farcs9$ in diameter, surrounded by a massive gas and dust ring,
which is responsible for most of the far infrared flux.  Subsequently,
\citet{forbes94} analysed 3 and 6~cm radio maps, combined with
\hal\ images and found a starburst ring of diameter $9\arcsec \times
7\arcsec$ ($0.86\mbox{\,kpc}\times 0.67\mbox{\,kpc}$).  For
comparison, the sample of 22 nuclear rings of \citet{mazzuca08} shows
diameters ranging from $0.4\mbox{\,kpc}\times 0.4\mbox{\,kpc}$ to
$3.4\mbox{\,kpc}\times 2.0\mbox{\,kpc}$.  The ring in NGC\,7552 looks
almost circular in projection, due to its inclination of only
$\sim28$\degr\ \citep{feinstein90}, and has a width of $3\arcsec -
4\arcsec$.  It contains about 60\% of the 6~cm radio flux.
\citet{forbes94} state that this ring cannot be seen in single colour
optical images largely because of dust extinction.  High resolution
NIR images taken with the SHARP~1 camera \citep{schinnerer97} showed
an inner ring-like structure with a diameter of $7\arcsec - 8\farcs5$,
and similar structures have been seen at MIR wavelengths
\citep{schinnerer97,siebenmorgen04,vanderwerf06}.

 \begin{figure}[ht]
    \centering
    \resizebox{\hsize}{!}{\includegraphics{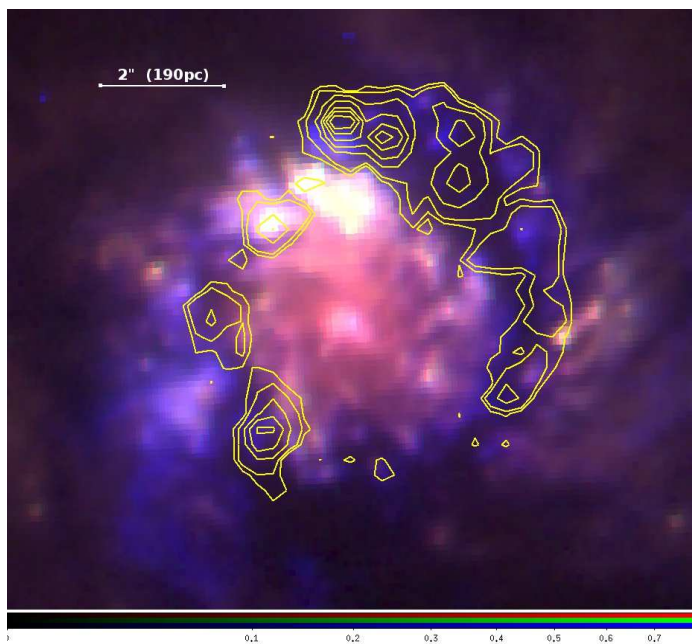}}
    \caption{Colour composite of the nuclear region of NGC 7552 made
      from archival HST/WFPC2 images in the F658N (\hal, blue) F336W
      (green), and F439W (red) filter.  The yellow contours outline
      the \NeII 12.8\m\ emission as detected with VISIR
      (Fig.\ref{fig:fourplots}\,a), the scale bar indicates $2''$ or
      190\,pc. North is up and East to the left.}
    \label{fig:whereclusters}
  \end{figure}

In order to assess the correspondence between the morphology seen in
our MIR observations and other wavelengths at highest angular
resolution, we have created a false colour image from archival
HST/WFPC2 data in the F336W and F439W broadband filters and the
\hal\ (F658N) narrowband filter.  Fig.~\ref{fig:whereclusters} shows
the central region, together with the contours of the VISIR
\NeII\ 12.8\m\ emission.  We find an excellent correspondence
between the heavily dust extincted ``dark ring'' and the
\NeII\ emission.  The HST images also confirm the early results of
\citet{feinstein90} of a central ($\leq 4\farcs9$) region of low
extinction surrounded by a massive ring of gas and dust.  However we
emphasize two findings:

First, the optically dark ring is the origin of most of the MIR
emission (Fig.~\ref{fig:fourplots}), and hence, the most recent
population of massive stars in the center of NGC\,7552 is still hidden
at optical wavelengths.  We also note that this happens on large
scales of hundreds of parsecs, not just locally for an individual
cluster.  Furthermore, the \brg\ map (Fig.~\ref{fig:fourplots}\,d)
resembles the MIR \NeII\ map (Fig.~\ref{fig:fourplots}\,a) very well.

Second, the reconstructed K~band continuum map
(Fig.~\ref{fig:fourplots}\,b), which is very similar to the broadband
image of \citet[][Fig.~2]{schinnerer97}, does {\em not} resemble the
\brg\ map at all, but looks smilar to the broadband optical colours
seen by HST (Fig.~\ref{fig:whereclusters}).  The northern K~band
continuum peaks (which must not be confused with the MIR peaks M1 \&
M3, which are at larger radii from the nucleus) coincide with the
bright regions in the HST image.  Most of the K~band continuum
emission comes from within the inner $3''$.  

As far as the structure within the ring is concerned we also observe
significant changes with wavelength.  The IRAC images
(Fig.~\ref{fig:irac}) show two peaks, the northern one, which is close
to our peaks M1 and M2, and the southern one, which coincides with the
position of M7.  The two peaks have similar brightness at 8\m, while
the southern one becomes the dominant peak at shorter wavelengths.
Our strongest 12\m\ source, M1, does not show significant emission in
the K~band continuum, whereas the nucleus appears as the brightest
peak in the K~band continuum image but is not detected at MIR
wavelengths.  A very interesting peak is M7 to the Southeast, which is
further discussed in section~\ref{sect:peak7}


\subsection{SEDs of Clusters and Ring}

Figure~\ref{fig:sed} displays the various mid-IR spectra and
photometric measurements to construct the spectral energy distribution
(SED) of the starburst ring of NGC\,7552.  Here we compare our VISIR
measurements with the literature data from Spitzer (IRAC, IRS) and ISO
(SWS, PHOT).  The ISOPHOT spectrum \citep{siebenmorgen04} was obtained
through a 24\arcsec\ aperture, and the \spitzer-IRS spectrum from the
SINGS Fifth Data Release \citep{kennicutt03} integrated over a
$27''\times 43''$ wide spectral map.  In addition, ISO-SWS
observations \citep[][ not shown in Fig.~\ref{fig:sed}]{verma03}
revealed, within an aperture of 14\arcsec$\times$20\arcsec,
\ArIII\ and \SIV\ line fluxes of $25\times10^{-21}$~W\,cm$^{-2}$ and
$3\times10^{-21}$~W\,cm$^{-2}$, respectively.

Within the given uncertainties, the agreement between the measured
fluxes from different instruments and observing modes is very good.
This may be surprising since the ISO and Spitzer measurements also
include the nucleus of NGC\,7552 whereas the ground-based measurements
focus on the ring only.  However, the contribution of the nucleus to
the total emission from the central region becomes increasingly less
toward longer wavelengths (cf. Figs. \ref{fig:fourplots}\,b and
\ref{fig:irac}) and is negligible at mid-IR wavelengths.
Furthermore, the fact that the \spitzer-IRAC photometry agrees so well
with the ISOPHOT measurements, and even the IRAS 12\m\ flux density
(Fig.~\ref{fig:sed}), suggests that essentially all of the mid-IR
emission comes from the central 12\arcsec\ region.  For the total
infrared luminosities of the individual clusters we refer to
section~\ref{sect:lumis}.


\subsection{Clusters versus Extended Emission}
\label{sect:extended}

Comparing the emission from the nine brightest mid-IR peaks to the
total flux integrated over the ring, however, we find a significant
mismatch.  The large ratios between compact to total (compact plus
diffuse) emission, quantified in Table~\ref{table:perc}, indicate that
there must be substantial diffuse emission within the ring and outside
the mid-IR peaks ($\Sigma$M$_i$).  The \NeII\ and 12\m\ continuum
flux densities emitted by all identified clusters are only 32\% and
25\%, respectively, of the emission integrated over the ring area.
The contribution of the clusters to the total emission in the
8.6\m\ PAH band is even less, only 5\%.  Moreover, most of these 5\%
has its origin in the mid-IR source M7 (with less than 1\%
contribution from the most luminous peak M1).  A substantial
contribution of diffuse emission of \NeII\ has also been observed
in the dwarf starburst galaxy NGC\,5253, where $\sim$ 80\% of the
\NeII\ line flux is diffuse and not directly associated with the
central super star cluster (SSC) \citep{martin:ngc5253,beirao06}.

\begin{table}
\caption{MIR emission from clusters versus total emission in the ring.}
\label{table:perc}
\begin{center}
\begin{tabular}{lccc}
\hline\hline
 & clusters\,/\,ring & clusters\,/\,total & ring\,/\,total\\
\hline\\[-5pt]

\NeII       & 0.32$^a$ & 0.15$^b$ & 0.48$^b$\\
12 \m\ cont.& 0.25$^a$ & 0.22$^b$ & 0.86$^b$\\
8.6 PAH     & 0.05$^a$ & 0.03$^c$ & 0.65$^c$\\

\hline
\end{tabular}
\end{center}

$^a$this work only;
$^b$incl.\ SINGS 5$^{th}$ data release;
$^c$incl.\ data from \citet{siebenmorgen04}.

\end{table}

We emphasise that ground-based observations in the mid-IR generally
possess a much reduced sensitivity to low surface brightness.  Here we
use the measured \NeII\ line fluxes for comparison:
\citet{siebenmorgen04} derived $(490\pm 56)\times
10^{-21}$\,W\,cm$^{-2}$ from TIMMI\,2 observations, \citet{verma03}
measured $680 \times 10^{-21}$\,W\,cm$^{-2}$ from ISO-SWS data, and we
derived $(800\pm 50)\times 10^{-21}$\,W\,cm$^{-2}$ for Spitzer-IRS from
the $5^{th}$ data release of the SINGS Legacy Team.  While the
\NeII\ line fluxes from VISIR (Table~\ref{table:big}) and TIMMI\,2 agree
reasonably well, the space-based measurements are typically about a
factor of two higher.  In other words, the ground-based measurements
underestimate the emission from resolved low surface brightness
regions.  This technical shortcoming may be even more relevant for the
derived PAH ratios.  Hence, the above mentioned ratios from the first
data column in Table~\ref{table:perc} should only be considered an
upper limit.

In summary, a large fraction of the \NeII\ and PAH band emission is
not directly associated with massive, young star clusters.  In
principle, the diffuse component may be due to an older population of
stars (e.g., an aged generation of previous super star clusters), a
more distributed mode of star formation (e.g., stars not localised in
super star clusters), or a large fraction of radiation from the
identified mid-IR peaks leaking out far into the surrounding
interstellar medium.  In any case, the localised emission diagnostics
do not exclusively trace the most recent sites of star formation and
need to be interpreted with care (see sections~\ref{sect:resolved},
\ref{sect:ages}).


\subsection{PAH Emission}
\label{sec:PAH}

The spectra in Fig.~\ref{fig:sed} are dominated by strong PAH emission
features at 6.2, 7.7, 8.6, 11.3 and 12.7\m. PAHs are considered the
most efficient species for stochastic, photoelectric heating by UV
photons in PDRs \citep{bakes94}.  They are commonly found in star
forming galaxies \citep[e.g.,][]{genzel00,smith07}.  However, intense
UV fields can also lead to the gradual destruction of PAH molecules
\citep[e.g.,][]{geballe89,cesarsky96,beirao06}.  PAHs may be
considered as overall good tracers of starburst activity in a
statistical sense \citep[e.g.][and references therein]{brandl06}.

The low-resolution VISIR N~band spectrum of peak M1 obtained is
shown in Fig.~\ref{fig:sed}.  It is characterised by a rising
continuum, typical for thermal emission by dust, a strong \NeII\ line,
and a weak 11.3\m\ PAH. We emphasise that the spectral flux density
agrees very well with the photometric fluxes obtained through the
VISIR filters, which agree well with the flux levels of the ISO and
\spitzer\ spectra.  Hence, we have good confidence in the absolute
flux calibration of the VISIR spectra. Nevertheless, the weakness of
the PAH emission feature in the spectrum is striking (see
section~\ref{sect:resolved}).

Using PAHs as quantitative tracers of star formation is known to have
some shortcomings.  First, at high angular resolution, the
correspondence between observed PAH strength and the luminosity of the
young clusters may break down, for reasons discussed in
section~\ref{sect:resolved}.  Secondly, PAHs may also be excited in
less UV-rich environments and can thus trace other sources besides
massive young stars, such as planetary nebulae and reflection nebulae
\citep[e.g.][]{uchida98,li02}, and B~stars \citep{peeters04}.  Since
these sources dominate the galactic stellar budget, i.e. are also much
more numerous in the volume that comprises the starburst ring in
projection, we expect significant emission from these sources, in
particular if the starburst has already been going on for while and we
do not observe the first generation of super star clusters any more.

Table~\ref{table:pahs} lists the continuum subtracted PAH band fluxes
for some mid-IR peaks and the integrated emission from the starburst
ring.  For comparison, measurements from TIMMI\,2, ISOPHOT and
\spitzer-IRS are also included.

\begin{table}
\caption{Continuum subtracted 8.6 and 11.3\,\m\ PAH band fluxes in units of $10^{-21}$~W\,cm$^{-2}$.}
\label{table:pahs}
\begin{center}
\begin{tabular}{lccc}
\hline\hline
Source & 8.6 PAH & 11.3 PAH & Aperture\\
\hline\\[-5pt]
M1            & $\sim0$ & 18$^{a}$   & 0\farcs7\\
M2            & $\sim0$ & --   & 0\farcs7\\
M7            & $61$    & --   & 0\farcs7\\
$\Sigma$M$_i$ & $\sim60$    & --   & 0\farcs7\\
Ring          & 1130    & --   & 3\arcsec--8\arcsec\\[2pt]
\hline\\[-7pt]
TIMMI2        & --      & 330  & 3\arcsec\\
ISOPHOT       & 1740    & 1620 & 24\arcsec\\
IRS/\spitzer  & --      & 2570 & 27\arcsec$\times$43\arcsec\\
\hline
\end{tabular}
\end{center}
$^{a}$ Measurement obtained from the VISIR spectrum of M1.
\end{table}


\subsection{Atomic Hydrogen Lines and Stellar Ages}
\label{sec:hydrogenlines}

The information provided by the atomic hydrogen lines listed in
Table~\ref{table:big} can be used in various ways.  The intensity of
the extinction corrected \brg\ line (section~\ref{sec:extinction})
can be used to calculate \citep[see e.g.,][]{ho90} the number of
hydrogen-ionising Lyman-continuum photons per second:
\begin{equation}
   N_{\rm Lyc} = 8.7\times 10^{45}\left( \frac{D}{\rm kpc} \right)^2
   \left( \frac{S_{{\rm Br}\gamma}}{10^{-12} {\,\rm erg\, cm}^{-2}\,
       {\rm s}^{-1}} \right) {\rm s}^{-1}.
\end{equation}

From $N_{\rm Lyc}$ we can estimate an equivalent number of O7V stars
($N_{\rm O7V}$).  We use the calibration of O star parameters provided
by \citet[][Table 4]{martins05} for the observational $T_{\rm eff}$
scale and a number of $\log (N_{\rm O7V}) = 48.75$ Lyman
continuum photons per O7V star.  Both $N_{\rm Lyc}$ and $N_{\rm O7V}$
are listed in Table~\ref{table:big}.

The equivalent widths of the hydrogen recombination lines H$\alpha$
and \brg\ are commonly used as age estimators of young stellar
populations.  These lines are predominantly produced by the most
massive stars, which are short-lived.  Hence, the strength of the
line-to-continuum ratio is a strong function of cluster age as the
most massive stars evolve.  

The evolution of the \brg\ EW with time has been modelled in detail,
e.g. by \citet[][ Starbust99]{leitherer99}.  However, the relatively
large distance to NGC\,7552, the low EW of \brg, and the dense ISM in
the starburst ring suggest that modelling the MIR peaks as isolated
young clusters may be insufficient.  We have therefore included
another step based on the combination of Starburst99 with the
photo-ionisation code Mappings \citep[e.g.][]{dopita00}, which
includes the radiation transfer in an evolving \HII-region.
\citet{groves08} have published a set of comprehensive, panchromatic
starburst models, from where we took the equivalent width of \brg\ as
a function of the cluster age.  We assume an instantaneous star
formation, a thermal pressure $\log(P/k) = 5$, and a so-called
compactness $\log{\cal C} = 5$.  This compactness parameter describes
the time evolution of the \HII-region, and is defined as
$\log{\cal C} \equiv \frac{3}{5}\log\left( \frac{M_{cluster}}{M_{\odot}} \right)
 + \frac{2}{5}\log\left( \frac{P/k}{{\rm cm}^{-3} {\rm K}} \right)$ 
(see \citet{groves08} for details).  

To derive the cluster ages we compared the extinction corrected EWs of
\brg\ (Table~\ref{table:big}) with the ages from the model.  The
resulting cluster ages are given in in the last column of
Table~\ref{table:big}.  

A couple of items are noteworthy considering our age estimates.
First, our approach is similar to the computed model tracks by
\citet[][Fig.~14]{snijders07}.  Second, in comparison to ``pure''
Starburst99 estimates, we derive cluster ages that are systematically
younger but only by $0.1 - 0.3$~Myr.  Third, the EW of \brg\ is only
accurate for young ages.  Beyond 6.5~Myr, EW(\brg) becomes very small,
observational and model uncertainties begin to dominate, and for
clusters older than 8~Myr \brg\ is even observed in absorption and may
reduce the emission features of nearby, younger clusters.  Finally,
the {\em relative} differences among our derived cluster ages,
however, can be considered quite accurate: the difference between a
cluster age of 5.5 and 5.8~Myr corresponds to a difference in EWs of
18\AA\ and 8\AA, respectively, which is much larger than the
spectrophotometric errors.
We shy away from quantifying formal errors on the individual age
estimates because we believe that the largest uncertainty may be given
by the assumption that each photometric aperture includes only one,
coeval stellar population.


\subsection{Extinction}
\label{sec:extinction}

The N~band includes the characteristic Si=O stretching resonance of
silicate-based dust, centred at 9.7\m. However, since it is usually
surrounded by strong PAH emission features to both sides (see
Fig.~\ref{fig:sed}, left) its utilization as a quantitative measure of
extinction in moderately dust enshrouded systems is very uncertain.
On the other hand, the \brg/Pa$\beta$ line ratio from the the SINFONI
IFU data allows us to create an extinction map by comparison with the
case B theoretical ratio of 5.87 \citep[assuming $T_e=10^4$~K and
  $n_e=100$~cm$^{-3}$,][]{hummer87}.  Assuming a near-IR extinction law
$A_\lambda \propto \lambda^{-1.8}$ \citep{martin90} and $R_V=3.1$
\citep{mathis90} we calculated $A_V$ and list the values for the
mid-IR peaks in Table~\ref{table:big}.  The average extinction of the
nine peaks is $A_V = 7.4$ and reaches from 5.2 to values as high as 9.0
for source M3.  Integrated over the entire ring area, the
average extinction reaches ``only'' $A_V = 5.3$.  

\begin{figure}[ht]
    \centering
    \resizebox{\hsize}{!}{\includegraphics{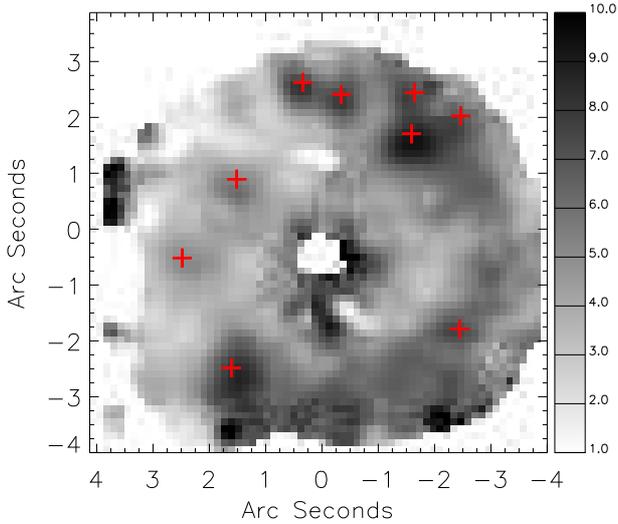}}
    \caption{Extinction map computed from the \pab\ and \brg\ line
      emission maps. The mid-IR peaks from Fig.~\ref{fig:fourplots}\,a
      are indicated by red crosses.}
    \label{fig:av}
  \end{figure}

With this information we constructed an extinction map, pixel-by-pixel
for the central region of NGC\,7552, which is shown in
Fig.~\ref{fig:av}.  These spatially resolved maps can be used to
correct the observed \brg\ line fluxes for extinction.  The extinction
corrected \brg\ line fluxes are listed in the third data column of
Table~\ref{table:big}.  We note that there is, generally, a very good
correspondence between the extinction peaks and our cluster positions.

\section{Results and Discussion}
\label{discussion}

Generally speaking, there are four main reasons why MIR
images of starburst regions may reveal a different morphology than
those taken at visible or near-IR wavelengths.  All these effects
should be kept in mind when discussing the properties of the ``MIR
clusters'' to characterise the starburst ring.

The first one is simply angular resolution which is proportional to
$\propto \lambda / D$. Since the wavelength $\lambda$ is typically
$5-10$ times longer and the telescope diameter $D$ is often smaller,
at least for space telescopes, the resulting angular resolution of the
MIR images is much reduced.  However, this effect does not play a role
in this analysis -- with the exception of the IRAC images presented in
Fig.~\ref{fig:irac} -- based on MIR images of $0\farcs3 -
0\farcs4$ resolution.
The second one is simply the generally much lower sensitivity of
ground-based MIR instruments.  This effect will introduce a selection
bias toward the most luminous clusters and overlook less luminous and
older, more evolved clusters.
The third one is dependency of the dust extinction on wavelength,
which may reveal heavily embedded clusters at MIR wavelengths while
they are invisible at shorter wavelengths.
The fourth effect is the fundamental difference in the physical
emission mechanism.  While visible and NIR images mainly reveal the
starlight of the cluster stars, the MIR emission is generally
re-radiated light from the dust around the cluster.  Fluctuations in
the surrounding dust distribution may thus affect the location
of the MIR peak, and substructures of smaller clusters, surrounded by
dust lanes may appear as one large complex.
In the following discussion we need to keep these issues in mind.


\subsection{Resolving the \HII\ Region Structure}
\label{sect:resolved}

In the most simplistic picture, the massive stars in a young cluster
provide a sufficient number of photons with $E_{\gamma} \geq 13.6$\,eV
to ionise the surrounding hydrogen.  The energetic radiation creates a
so-called \HII\ region from where most of the emission from the ionic
states with higher excitation potential originates:
\NeIII\,(40.96\,eV), \NeII\,(21.56\,eV), \SIV\,(34.97\,eV),
\SIII\,(23.34\,eV), and \ArIII\,(27.63\,eV).  The \HII\ region is
surrounded by molecular gas which is ``only'' exposed to far-UV
radiation ($6 - 13.6$\,eV), which strongly influences its chemical and
thermal structure \citep{tielens85}, and which is responsible for most
of the PAH emission.  Beyond this photo-dissociation region (PDR) we
expect the diffuse and partially neutral interstellar medium (ISM).
In the most simplistic picture, the size of the \HII\ region is given
by the Str\"omgren radius
\begin{equation}
R_S = \left(\frac{3}{4 \pi} \frac{N_{\rm Lyc}}{n^2 \beta_{\rm B}} \right)^{1/3}.
\end{equation}

If we assume a density of $10^{3}$\,cm$^{-3}$, a recombination
coefficient $\beta_{\rm B} = 2.6\times 10^{-13}$\,cm$^{3}$/s, and a
Lyman continuum photon flux of $N_{\rm Lyc} = 5\times
  10^{51}$\,s$^{-1}$ (Table~\ref{table:big}) we get $R_S = 5.4$\,pc or
  a region of about 10.8\,pc in diameter in projection.

In reality, however, the structure of giant PDR/\HII\ regions is
significantly more complex as illustrated by the closer examples of
30\,Doradus in the LMC \citep[e.g.,][Fig.\,1]{indebetouw09} and
NGC\,604 in M\,33 \citep[e.g.,][Figs.\,2\,\&\,3]{hunter96}.  The main
differences between the Str\"omgren picture and these giant
\HII\ regions are many-fold: they are not powered by a single,
point-like cluster, but by several, distributed clusters; dust competes
with the gas for ionising photons and will shrink the size of the
\HII\ region; the radiation will not radiate perfectly isotropic and
the interfaces between \HII\ region, and PDR are also shaped by the
stellar winds, outflows and supernovae.  Hence, the above estimated
$R_S$ may only provide a rough estimate. In reality, the luminous PDRs
are likely at larger distance from the centre of the \HII\ region.  We
would thus expect that spectra of the central cluster, taken at high
angular resolution, will miss large contributions from the PDR.

At the given distance to NGC\,7552 of 19.5~Mpc and a slit width of
0\farcs75, the VISIR spectra cover a region of approximately 70~pc in
size.  Indeed, the discrepancy in the strength of the 11.3\m\ PAH
feature between the high resolution VISIR spectrum (black line in
Fig.~\ref{fig:sed}) and the spatially integrated (and rescaled)
\spitzer\ spectrum (red line in Fig.~\ref{fig:sed}) is striking.
Similarly, the ratios in Table~\ref{table:perc} are based on
photometric aperture sizes of 65~pc.  While, within the same aperture
size, the clusters contain 32\% of the \NeII\ emission of the ring,
they contain only 5\% of the PAH emission.  In other words, zooming in
on radial distances of 33\,pc around the clusters reduces the observed
PAH strength by a factor of six.  This is a clear indication that we
are resolving the \HII/PDR complex.

This may not come as a big surprise.  The strong dependency of mid-IR
spectral diagnostics on the aperture size has already been pointed out
by \citet{martin06}, and a similar effect was noted for the young
super star clusters in the overlap region of the Antennae galaxies by
\citet{snijders07,brandl09}.  However, we want to emphasise two
findings: First, a significant fraction of the \NeII\ emission escapes
from the cluster regions.  Second, at high angular resolution, the
mid-IR spectra of the central clusters include only partial
contributions from the PDR.

Both findings have important general implications: On one hand, the
physical properties of the young stellar populations (such as age or
initial mass function, IMF) derived from lines fluxes integrated over
the entire galaxy are very uncertain, as a large fraction of the line
flux may not be associated with those populations.  On the other hand,
comparing spectral features that originate from physically and
spatially different regions (such as ionic lines from \HII\ regions
and PAHs from PDRs) only for small apertures may lead to similar
systematic errors.


\subsection{Cluster Luminosities and Star Formation Rates}
\label{sect:lumis}

An important quantity to characterise starbursts is their bolometric
luminosity.  In very dusty environments ($A_V \gg 1$) we assume that
the bolometric luminosity $L_{\rm bol}$ is approximately given by the
total (8 -- 1000\m) infrared luminosity $L_{\rm IR}$.  The latter
is commonly derived \citep[e.g.,][]{sanders03} from the four IRAS
bands using the equation $L_{IR} = 562860\cdot D^2\cdot (13.48S_{12\mu
  m} + 5.16S_{25\mu m} + 2.58S_{60\mu m} + S_{100\mu m})$, where
$S_{\lambda}$ is in Jy and $D$ in Mpc.  From the spatially integrated
IRAS fluxes of NGC\,7552 in we get $L_{IR} = 8.5\times 10^{10}$\lsun.
Considering only the starburst ring, \citet{schinnerer97} quote a
bolometric luminosity of $L_{bol} = 2.8\times 10^{10}$\lsun.

Of particular interest are the infrared luminosities of the individual
clusters.  However, observations that cover the FIR regime do not
possess the required spatial resolution, and the VISIR observations
only cover 8 -- 13\m.  Hence, we bootstrap the VISIR fluxe densities
measured through the `NeII\_2' filter (Table~\ref{table:fluxes}) to
$L_{\rm IR}$ estimates of clusters from the literature.  The `NeII\_2'
filter has been chosen as the longest wavelength filter of our data
set, which does not not include significant emission or absorption
features.  \citet[][, Table 9]{brandl09} have provided SEDs and total
infrared luminosities for six massive clusters in the Antennae
galaxies.  We have computed the ratios between the `NeII\_2' filter
band averaged $F_{12\mu m}$ and $L_{\rm IR}$ for these six clusters.
Adjusted for the distance of NGC\,7552 we get the empirical relation
\begin{equation}
L_{\rm IR} = (30.1\pm 7.9)\times F_{12\mu m}, 
\end{equation}
where the 12\m\ flux density is given in mJy and $L_{\rm IR}$ in
units of $10^6$\lsun.  The quoted uncertainty is the standard
deviation of the six derived ratios.

The estimated infrared luminosities are listed in
Table~\ref{table:big}.  More than half of the total luminosity is
provided by the clusters M1, M2 and M7.  Summing up the total
luminosity of all MIR peak yields $L_{IR}^{\rm all} = 2.1\times
10^{10}$\lsun.  This is 75\% of the bolometric luminosity of the
starburst ring derived by \citet{schinnerer97}.  The good agreement of
these two estimates, derived by completely different methods, supports
our ``bootstrapping'' approach.

\citet{kennicutt98} has shown that the star formation rate (SFR) can
be derived from $L_{\rm IR}$ via: SFR\,[\msun\,yr$^{-1}] = 4.5\times
10^{-44} L_{\rm IR}$\,[erg s$^{-1}$].  We note that this conversion
strictly applies only to the continuous star formation approximation
with ages of order $10 − 100$\,Myr. Since our clusters are younger,
the IR luminosity per unit mass of stars formed will be somewhat
lower, and the \citet{kennicutt98} relation will overestimate the true
SFRs.  (This limitation would also apply to the common method of
directly estimating the SFR from the IRAS fluxes).  With the above
conversion $L_{IR}^{\rm all} \approx 2.1\times
10^{10}$\lsun\ corresponds to a SFR of 3.7\msun\,yr$^{-1}$ within only
the main clusters in the starburst ring.  For comparison,
\citet{schinnerer97} derived from the radio luminosity of the radio
knots a star formation rate of 1\msun\,yr$^{-1}$ in the ring of
NGC\,7552, and \citet{calzetti10} derived from Spitzer photometry
taken by the SINGS team 10.3\msun\,yr$^{-1}$ for the entire galaxy.
The sample of 22 nuclear ring galaxies of \citet{mazzuca08} covered
SFRs from 0.1\msun\,yr$^{-1}$ to 9.9\msun\,yr$^{-1}$, with a median
value of 2.2\msun\,yr$^{-1}$.  Our value of 3.7\msun\,yr$^{-1}$ fits
very well within these numbers.
%


\subsection{Cluster Ages}
\label{sect:ages}

The ages of clusters M1 -- M9 are listed in the last column of
Table~\ref{table:big} and lie between 5.5 and 6.3~Myr with a mean age
of $5.9\pm 0.3$~Myr.  Although they have been derived under the
simplifying assumption of instantaneous star formation we can draw two
important conclusions:
First, there are {\em no} luminous young clusters associated with the
mid-IR peaks that indicate very recent ($\ll 5$~Myr) star formation.
Second, the age spread between the clusters associated with the
mid-IR peaks is relatively small.

To investigate whether high extinction may hide the youngest clusters
even at K~band, we also use independent diagnostics at MIR
wavelengths.  A commonly used age diagnostic is the ratio of the two
neon fine structure lines, \NeIII\ and \NeII, because it is largely
independent of density and extinction effects.  With excitation
potentials of 40.96\,eV and 21.56\,eV, respectively, the ratio of
\NeIII\,/\,\NeII\ measures, to first degree, the hardness of the
radiation field, which is mainly dominated by the massive O~stars and
thus a strong function of age and initial stellar mass function (IMF)
of an instantaneous starburst.  (We note that factors other than age
may also play a role in affecting the observed line ratios, such as
high metallicity (via stellar evolution and line blanketing), high
electron density \citep{snijders07}, and variations of the upper IMF).
Unfortunately, we have no measure of the \NeIII15.56\m\ line, which
lies outside the atmospheric transmission window, but we can still
estimate the MIR fine structure line ratios in two different ways:

First, \NeIII\ has been measured from space.  Both ISO-SWS and
Spitzer-IRS found rather low \NeIII\,/\,\NeII\,$\approx 0.08$ through
large apertures that covered the central region.  The large
discrepancy between compact and extended emission, as discussed in
section~\ref{sect:extended}, complicates the interpretation of the
line ratios.  However, if we assume that most of the \NeIII\ is only
produced by the massive O~stars in the clusters -- which is not an
unrealistic assumption given the high excitation potential of
\NeIII\ -- we can compute the ratio between the \NeIII\ flux given by
\citet{thornley00} and the sum of \NeII\ from all clusters
(Table~\ref{table:big}).  Under this assumption, we derive
\NeIII\,/\,\NeII\,$\approx 0.41$.  Starburst models
\citep[e.g.][]{snijders07} indicate that a ratio of 0.4 is in good
agreement with an age of approximately 5~Myr.  


Second, for the brightest MIR peak M1 we obtained a VISIR spectrum
that covers also the \SIV10.51\m\ line.  Unfortunately, the \SIV\ line
was not detected in the VISIR spectrum of M1, so we need to follow a
different route.  
\SIV\ has an excitation potential of 34.97\,eV, which is close to the
40.96\,eV of \NeIII, and may thus serve as a substitute.
\citet[][]{groves08b} provide an empirical calibration of
\SIV\,/\,\NeII\ against \NeIII\,/\,\NeII, parametrised as:
\begin{equation}
\log\left( \frac{\NeIII}{\NeII} \right) = \alpha \log\left(
\frac{\SIV}{\NeII} \right) + \beta .
\end{equation}
From fits to a large sample of 97 extragalactic \HII\ regions and 56
starburst galaxies \citet[][]{groves08b} derived the parameters
$\alpha = 0.82\ (0.65)$ and $\beta = 0.36\ (0.32)$ for extragalactic
\HII\ regions (starburst galaxies).  
From the \NeII\ flux listed in Table~\ref{table:big}, the above
correlation(s), and \NeIII\,/\,\NeII\,$\approx 0.41$, we would expect
a \SIV\ line flux from peak M1 of $F_{\SIV}^{\HII} = 3.1\times
10^{-21}$\,W\,cm$^{-2}$ ($F_{\SIV}^{\rm SBs} = 2.1\times
10^{-21}$\,W\,cm$^{-2}$).  In both cases, the line is too weak to be
detected in our VISIR observations.  Of course, this estimate assumes
that the properties of M1 are similar to those of the other clusters
in the starburst ring.

Independently, we can derive an upper limit on the \SIV\ line flux
from the noise in the 10.4 -- 10.6\m\ range.  This yields a $3\sigma$
upper limit of $F_{\SIV}^{\rm UL} = 2.3\times 10^{-20}$\,W\,cm$^{-2}$.
Hence, we derive an upper limit on the ratio of $F_{\SIV}^{\rm UL} /
F_{\NeII} \leq 0.12$.  Comparing this ratio with the models of
\citet[][ Fig.\,9]{snijders07} yields a lower limit of 4\,Myr on the
age of peak M1, consistent with the \brg\ age estimate.


\subsection{The Intriguing Peak M7}
\label{sect:peak7}

A rather unusual and thus very interesting source is peak M7, located
to the Southeast of the center.  A qualitative comparison between the
three prominent peaks, M1, M7, and the nucleus, is given below, where
`+' means bright and `-' means faint or undetected:

\begin{center}
\begin{tabular}{ l c c c }
   & K$_{\rm cont}$ & 8.6\m\ PAH & \NeII\ \\ 
\hline
Nucleus & + & - & - \\
M1      & - & - & + \\
M7      & + & + & + \\
\end{tabular}
\end{center}

M7 is equally prominent in \brg, \NeII, and the K~band continuum.
What is most remarkable, however, is its strong 8.6\m\ PAH emission.
Fig~\ref{fig:raw} (center) shows that M7 is the brightest 8.6\m\ PAH
emitter in the center of NGC\,7552. Furthermore, M7 is a peculiar
source in two more regards.

First, the velocity field of the nuclear region (see
section~\ref{sfhring} and Fig.~\ref{fig:brgvelofield}), while overall
relatively symmetric, reveals a distinct ``nose'' in the contour line
of $+50$\,km\,s$^-1$, approximately 3\arcsec\ to the Southwest, where
M7 is located. Apparently, the velocity field in the region of M7 is
slightly disturbed.

Second, \citet{rosenberg12} have recently investigated the
relationship between the \FeII $1.26\mu$m luminosity and the supernova
rate in a sample of 11 nearby starburst galaxies, including NGC\,7552.
They derived the supernova rate from Starburst99 modelling of the
\brg\ emission on a pixel-by-pixel basis and found a very good
correlation across all 11 galaxies between the derived supernova rates
and the measured \FeII\ line luminosities -- with one exception, namely
M7 in NGC\,7552.  According to \citet[][their Fig.\,7]{rosenberg12},
the supernova rate derived from the \brg\ emission exceeds by far the
corresponding \FeII\ intensity.

We note that the derived age of M7, its dust extinction,
\brg\ equivalent width, and total luminosity are well within the range
of parameters of the other peaks (Table~\ref{table:big}).
Nevertheless, peak M7 is not likely a simple, single massive cluster
(see section~\ref{sect:clusterpeak} for a discussion of possible
causes.)


\subsection{Comparison with other Massive Clusters}
\label{clustercomparison}

The main question here is: Are the MIR selected clusters in NGC\,7552
in any way special, or similar to massive young clusters in other
starburst galaxies?  This can be best addressed by comparing the
cluster properties in Table~\ref{table:big} with several ``reference
clusters'' from the literature (Table~\ref{tbl:otherclusters}).

\begin{table*}
\caption{Data of selected luminous clusters from the literature for comparison}
\label{tbl:otherclusters}
\begin{center}
\begin{tabular}{l@{\hspace{7pt}}c@{\hspace{7pt}}c@{\hspace{7pt}}c@{\hspace{7pt}}c@{\hspace{7pt}}c@{\hspace{7pt}}c@{\hspace{7pt}}c@{\hspace{7pt}}c@{\hspace{7pt}}c@{\hspace{7pt}}c@{\hspace{7pt}}c@{\hspace{7pt}}c}
\hline\hline\\[-5pt]       
      & Galactic & Starburst & NGC\,3603$^{c}$ & NGC\,5253$^{d}$ & \multicolumn{2}{c}{NGC\,1365$^{e}$} & \multicolumn{2}{c}{NGC\,4038/39$^{f}$} & \multicolumn{2}{c}{Arp\,143$^{g}$} & \multicolumn{2}{c}{M\,82$^{h}$} \\ 
      & \HII\ regions$^{a}$ & galaxies$^{b}$ & & C2 & M\,5 & M\,6 & peak 1 & peak 5 & knot E & knot F & region 2 & region 3 \\ 
\hline\\[-5pt]
$L_{IR}$ [$10^{9} L_{\sun}$] &  & 50 & 0.01 & 1.0 &    & 0.8 & 11.4 & 3.9 &    &   &  &  \\
$M$ [$10^{6} M_{\sun}$]      &  &    &      & $\approx 1$    & 38 & 25  &  3   &     & 25 & 6 &  &  \\
$A_{V}$ [mag] &  & 4.3 & $\le 1$ & 17 & 3.2 & 8.5 & 4.5 & 10.5 & 1.2 & 1.3 & $\approx 15$ & $\approx 25$ \\
Age [Myr]     &  &     &  $\sim 1$    & $<6$ & 7   & 7   & 2.5 & 4.5 & 3.5 & 2.5 & 5.5 & 5.5 \\
\NeIII\ / \NeII & 0.8 & 0.29$^{i}$ & 3.13 & 8.4 &  &  & 0.63 & 0.69 & 0.52 & 0.67 & 0.13 & 0.19 \\[5pt]
\hline

\end{tabular}
\end{center}

$^{a}$\citet{giveon02}
$^{b}$\citet{thornley00,brandl06}
$^{c}$\citet{lebouteiller08}
$^{d}$\citet{martin:ngc5253}
$^{e}$\citet{galliano08}
$^{f}$\citet{brandl09}
$^{g}$\citet{beirao09}
$^{h}$\citet{beirao08}
$^{i}$mean of 20 starburst galaxies, excluding the dwarfs II\,Zw\,40 and NGC\,5253.

\end{table*}

While the globally integrated ratio of \NeIII\,/\,\NeII\,$\approx
0.08$ is significantly below the mean value of 20 starburst galaxies,
the ``corrected'' value of 0.41 (section~\ref{sect:ages}) places
NGC\,7552 well within the range of starbursts and only a factor of two
below the mean radiation hardness of Galactic \HII\ regions.  The real
outlier in this regard in Table~\ref{tbl:otherclusters} is NGC\,3603,
a very young Galactic \HII\ region, dominated by a single compact
cluster, and NGC\,5253.  For the latter it was shown that the bulk of
the high-excitation \SIV\ and \NeIII\ fine structure line emission is
also associated with the single compact cluster C2, whereas the
diffuse component -- showing \ArII, \SIII, and \NeII\ emission -- is
much more extended \citep{martin:ngc5253,beirao06}.  In terms of Lyman
continuum photons $N_{\rm Lyc}$, our peak M1 is only a factor three
below the radio super-nebula NGC\,5223~C2 \citep[$2\times
  10^{52}$\,s$^{-1}$;][]{martin:ngc5253}.  However, in terms of the
ratio between ionising photons and the surrounding PDR mass, the peaks
in NGC\,7552 resemble the massive clusters in more luminous starburst
galaxies, while NGC\,3603 and NGC\,5253 may be better compared to blue
compact dwarf galaxies.  

The total infrared luminosities of the peaks M1 through M9 are, on
average, somewhat fainter than the most luminous Antennae cluster, but
agree well with e.g., NGC\,5223~C2, NGC\,1365~M6 and
NGC\,4038/39~peak\,5, which display similarly high extinction.

Since we have no good direct tracer of stellar mass (e.g. dynamical
masses) we refrain from providing cluster mass estimates here.
However, a comparison with the reference clusters of similar ages,
$L_{\rm IR}$ and $N_{\rm Lyc}$ in Table~\ref{tbl:otherclusters}
suggests that our clusters have typical masses of a few million solar
masses, which is not extraordinarily large.  Significantly more
massive young clusters, up to $3 \times 10^7$\msun, have been found
e.g. in NGC\,1365 by \citet{galliano08}.

The ages of the clusters in NGC\,7552 are also well within the range
of the MIR luminous clusters in Table~\ref{tbl:otherclusters}.  It is
evident that most of the massive clusters listed here managed to
retain a substantial amount of extinction after $5-7$\,Myr.  An
exception here is Arp\,143.  We note, however, that the starburst ring
in Arp\,143 results from a shock wave triggered by the recent head-on
collision between the two galaxies, NGC\,2444 and NGC\,2445
\citep[e.g.,][and references therein]{beirao09}.  This scenario is
physically different from dynamical instabilities in a quasi-stable
nuclear region, and this difference will be reflected in the density
of the ISM and its gas supplies (see also
section~\ref{clusterevolve}).  In summary, we conclude that the
clusters in the ring of NGC\,7552 appear like typical unit cells of
starbursts.


\subsection{``Clusters'' and ``Peaks'' may be Misleading Terms}
\label{sect:clusterpeak}
 
So far we have, in a very simplistic way, equated a MIR ``peak'' with
a super star cluster.  Both species are closely related in starbursts,
but there may not necessarily be a direct correspondence.  In
particular, one MIR peak may consist of several smaller clusters.  We
remind the reader that our angular resolution at near- and mid-IR
wavelengths corresponds to about 30\,pc, which is about one to two
orders of magnitude larger than the typical core radius of a massive
cluster \citep[e.g., R136,][]{brandl96}.  On one hand, it is likely
that the hierarchical cloud fragmentation in a dense and turbulent ISM
will lead to a conglomerate of smaller clusters
\citep[e.g.,][]{bonnell03}.  On the other hand, once a massive cluster
has formed in a dense environment it may trigger the formation of
other clusters in its vicinity \citep[e.g.,][]{deharveng05}.  While
both scenarios impact dynamical mass estimates and cluster evolution,
the latter scenario has the most severe effect as the peak cannot be
any longer considered a coeval population, formed in an instantaneous
event.

Observationally, there is strong evidence for a more complex scenario.
For instance, \citet{bastian09} found that the youngest clusters in
the Antennae are not isolated but part of an extended hierarchy of
star-forming regions.  In those ``nuclei'' of star formation, one
cluster is often surrounded by other clusters, which cannot be
resolved with the current combination of telescopes and wavelengths.
In the overlap region of the Antennae galaxies, \citet{snijders07}
found on scales below the resolution limit a complex structure, which
contains several young stellar clusters embedded in clumpy gas.
Similarly, in a statistical analysis of Antennae clusters
\citet{mengel05} found that the 10~Myr old clusters with high
extinction are preferentially located near younger clusters.  This
suggests that in environments where the gas is not rapidly removed,
ongoing -- maybe even triggered -- cluster formation is likely to
happen.

These aspects are very relevant to the derived cluster properties as
an unresolved complex of (sub-)clusters with significant age gradient
will appear differently.  While the high excitation lines
(e.g. \NeIII) will be dominated by the young cluster, the lower
excitation lines (e.g. \NeII) will primarily originate in the older
components, lowering the combined neon line ratio.  In addition, the
IMF -- when normalised to the most luminous stars -- will appear
steeper (bottom heavy). Furthermore, dynamical mass estimates yield
incorrect results since all components of the cluster complex will
contribute to the diagnostic line but are not in virial equilibrium.
And finally, significant patches of dust may be trapped in lanes
between sub-clusters and remain associated with the observed peak for
a longer time, as discussed in the next section.


\subsection{Cluster Evolution and Extinction}
\label{clusterevolve}

Table~\ref{table:big} shows that the clusters of NGC\,7552 have an
average extinction of 7.4~mag (at an average age of 5.9~Myr).  The
large amount of gas and dust, still associated with our MIR peaks
after relatively long time, indicates that the gas and dust dispersion
has been rather inefficient during the first six million years.
However, selecting clusters from MIR data introduces a selection bias
toward the most heavily extincted clusters.  This can be best
illustrated for the Antennae galaxies:

The {\em optically selected} clusters within the Antennae have
extinctions of $\le 4$~mag (with a mean of 2.6~mag in the age range
0--4~Myr \citep{whitmore02}.  In a study of {\em K~band selected}
clusters in the Antennae \citep{mengel05} found strongly variable
extinction with an average value of $A_{\rm V}=1.3$~mag.  However, the
{\em MIR-selected} clusters in the dusty overlap region of the
Antennae have extinctions of $A_{\rm V} > 4$ up to $A_{\rm V} \sim
10$~mag \citep{brandl09}.  Although the latter may not dominanate by
numbers, they are no exotic species either: \citet{whitmore10} found
that $16\% \pm 6\%$ of the IR-bright clusters in the Antennae are
still heavily obscured with values of $A_{\rm V} >3$~mag.
\citet{mengel05} found a clear trend of lower $A_{\rm V}$ with
increasing age.  Several studies \citep[e.g.][]{mengel05,whitmore10}
suggest that the typical time scale for massive clusters to emerge
from their natal dust cocoons is less than 10\,Myr.

However, the local environment and the location of the clusters within
the galaxy is likely to play an important role in the efficiency of
the dust removal.  It is conceivable that the expansion of the
\HII\ region goes faster in lower density environments (e.g., in dwarf
galaxies or the outer spiral arms) than in higher density regions of
galactic nuclei where the gas and dust are trapped in a potential and
even get continuously replenished.  \citet{mengel05} argue that an
older cluster which suffers high extinction may no longer be directly
surrounded and obscured by its own dust cocoon, but just happens to be
located in a denser region of more recent star formation with higher
dust content.  

While this picture is likely to apply to clusters in starburst regions
in general, and to clusters in the starburst ring of NGC\,7552 in
particular, we would like to reemphasize the ``special'' and distinct
nature of MIR-selected ``peaks''.  Several studies \citep[e.g.,
][]{whitmore02,whitmore10} have found that, by number, the vast
majority of hundreds of clusters in the Antennae is optically visible
and can be well studied with the HST.  On the other hand, the few
highly reddened MIR peaks in the overlap region, although almost
insignificant by number counts in comparison, account for
approximately half of the total luminosity -- and thus of the total
star formation -- of the Antennae \citep{brandl09}.  It is likely that
the latter species does not exist in galactic disks and can only be
found in the densest environments.  However, the study of clusters in
starburst nuclei, ulta-luminous and sub-millimeter galaxies requires
infrared observations at high angular resolution
(cf. Fig.~\ref{fig:whereclusters}).


\subsection{Star Formation History in the Ring}
\label{sfhring}

In the Introduction (section~\ref{sect:intro}) we have summarised the
controversy around the physical origin and location of starburst
rings.  To complicate matters, the starburst ring does not appear to
be an narrow ring with a series of young clusters lined up on a
circle.  Instead, we observe (e.g., Fig.~\ref{fig:fourplots}) an annular
region of $\sim200$\,pc width with increased and non-uniform star
formation activity.

Here we take the existence of a ring for granted and want to
investigate how and where the clusters form within that ring.  We
assume that the gas gets efficiently transported from the outer
regions to the centre via the nuclear bar
\citep{schwarz81,ellison11}. The gas moves radially inward along the
bar dust lanes and accumulates at the so-called contact points, the
intersections between the dust lanes and the ring.  These contact
points are expected to be approximately perpendicular to the position
angle of the bar major axis \citep{regan03}.  Since the orientation of
the major axis of the bar is roughly East-West while the contact
points are to the North and South \citep{schinnerer97}, the
observations support the theory.  However, the details of where, how,
and when the gas gets turned into stellar clusters are still uncertain
and could possibly be explained with one of the following two scenarios.

In the first scenario, gas accumulates gradually in the nuclear region
along the ring.  Once a critical density is reached locally, the gas
becomes unstable to gravitational collapse, and shocks may trigger the
formation of a massive star cluster \citep{elmegreen94} or lead to
further cloud fragmentation due to supersonic turbulence
\citep[e.g.,][]{klessen01,bonnell03}, resulting in an ensemble of
clusters at that location.  Since these instabilities occur randomly
with respect to the galactic structure, no systematic age gradients or
patterns within the ring would exist.

In the second scenario, most of the gas enters the ring at the
intersection between the bar and the ring.  Shocks play an important
role here to get rid of the angular momentum, which is essential to
move the gas into the central ring \citep[see e.g.,][and references
  therein]{mazzuca08}.  The pile-up of gas at these two contact points
can be expected to ignite star formation in short-lived bursts. The
clusters formed here will orbit along the ring, and age as they travel
along the ring. In the meantime, new starburst events near the contact
points may produce new generations of young star clusters.  Hence,
observationally, one would expect to find the youngest clusters of the
ring near the contact points and increasingly older clusters in the
direction of rotation.

Some evidence for the latter scenario was provided by \citet{boker08}
for several galaxies, and in particular for NGC\,613.  In a more
comprehensive study of 22 barred galaxies , \citet{mazzuca08} found
that 10 out of 22 ring galaxies showed random variations in their
\HII\ region properties with no apparent age gradient.  However, 9
(out of 22) of their rings showed partial gradients along 25\% -- 30\%
of their rings.  Only three galaxies (NGC\,1343, NGC\,1530 and
NGC\,4321) showed a clear bipolar age gradient along the ring.
\citet{mazzuca08} could not relate the presence or absence of age
gradients with the morphology of the rings or their host galaxies.

 \begin{figure}[ht]
    \centering
    \resizebox{\hsize}{!}{\includegraphics{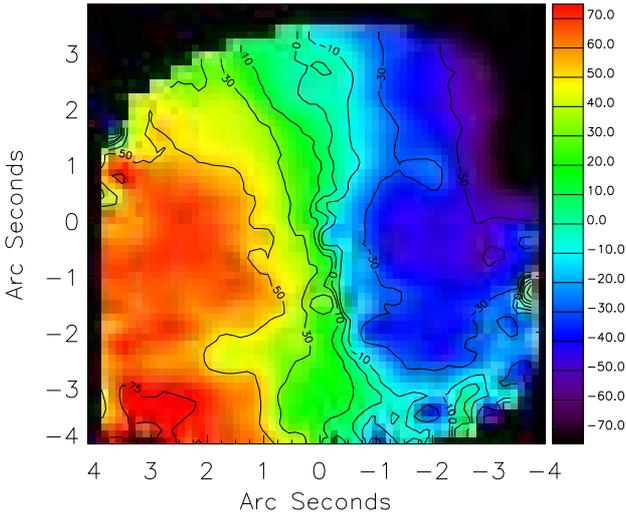}}
    \caption{Velocity field in [km/s] of the nuclear region of NGC
      7552 derived from the 2.166\m\ \brg\ line shifts.}
    \label{fig:brgvelofield}
  \end{figure}

In order to investigate a possible age gradient across the clusters in
NGC\,7552 we constructed a velocity map from the SINFONI \brg\ data
(Fig.~\ref{fig:brgvelofield}).  The line offsets were converted into
rotational velocities for a recessional velocity of NGC\,7552 of
$(1640 \pm 12)$\,km\,s$^{-1}$ \citep[][including ``Virgo+GA+Shaply'',
]{mould00}.  With this velocity offset the kinematic centre position
appears to lie close to the nucleus of NGC\,7552 as traced by the
K~band continuum peak, and the velocity field appears rather symmetric
to both sides of the centre.  The zero velocity line of the
\brg\ image is at the position angle of $\sim9$\degr\ with the
kinematic major axis at $\sim99$\degr.  The physical velocity of a
ring element is then related to the observed, Doppler-shifted velocity
via:
\begin{equation}
v_{ring} = \frac{v_{obs}}{\sin(i)\cos(\theta)},
\end{equation}
where $i$ is the inclination of 28\degr\ \citep{feinstein90} and
$\theta$ is the angle in the plane of the ring measured from the
position angle.  
%
%
From Fig.~\ref{fig:brgvelofield} we derive a mean $v_{ring} \approx
150$\,km\,s$^{-1}$, which is in perfect agreement with the typical
rotation velocity of $v_{\rm{rot}} = 150$\,km\,s$^{-1}$
\citet{mazzuca08} found for their sample of nuclear rings.

We can now estimate the travel time of a cluster at $r = 275$\,pc
along a quarter (90\degr) segment of the ring via:
\begin{equation}
t_{\rm travel} = \frac{l_{\rm segment}}{v_{\rm ring}} =
             \frac{\frac{\pi}{2}\times 275 {\rm pc}}{150 {\rm km\,/\,s}} \approx
             2.8\mbox{\,Myr}
\end{equation}

 \begin{figure}[ht]
    \centering
    \resizebox{0.85\hsize}{!}{\includegraphics{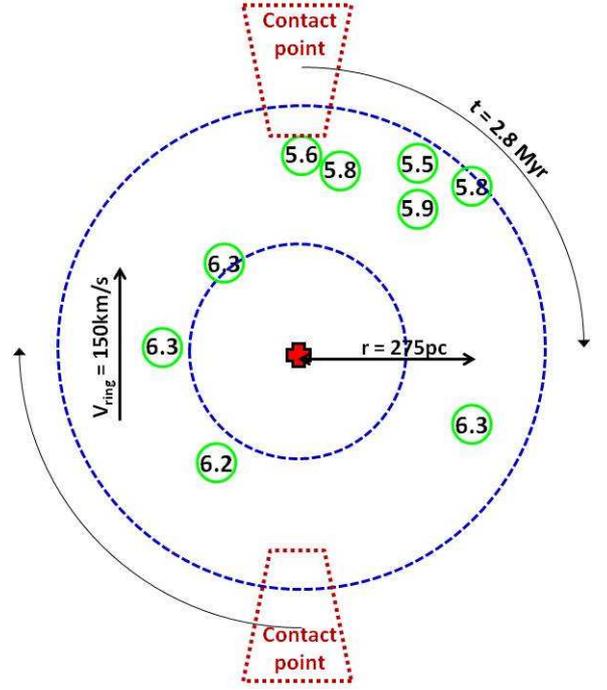}}
    \caption{Schematic outline of the cluster age distribution with
      respect to the location of the contact points in NGC\,7552.  The
      green circles indicate the positions of the MIR peaks and the
      inscribed numbers their ages in million years.  The direction
      and time scale of rotation are also indicated.}
    \label{fig:popcornring}
  \end{figure}

Fig.~\ref{fig:popcornring} shows a simplified representation of the
scenario for NGC\,7552 with both cluster locations and ages, and the
travel timescales indicated.  The youngest knots M1 and M4 appear
indeed to be located at or near the Northern contact point, and the
oldest knots M6, M8 and M9 are located more than 90\degr\ away from
the contact points.  This signature provides evidence that the cluster
age distribution is not completely random but somehow connected to the
contact points of the ring.  Our finding also agrees with the result
of \citet{mazzuca08} that in two-thirds of their 22 barred galaxies
the location of the {\em youngest} \HII\ region is within 20\degr\ of
the contact points.

Despite the limited number of MIR peaks, we find some weak evidence
for an age gradient in the direction of the ring rotation.  However,
the gradual age difference of the clusters is much smaller than the
estimated cluster travel time.  The presence of older clusters and the
lack of very young clusters close to the contact points raise some
doubts in this simple (second) scenario.  Our observations clearly do
not support the hypothesis that the high gas mass inflow rate will
immediately trigger gravitational collapse and ignite massive cluster
formation.

However, since the cluster ages are relatively large, they are likely
to have travelled a significant distance from their birthplace to
their observed location.  It is intriguing that the age of peak M1,
now located near the northern contact point, is exactly the travel
time from the southern contact point to its current location (half a
ring orbit).  It appears therefore possible that a dramatic episode of
gas inflow at the contact points about 5.6\,Myr ago has triggered
massive cluster formation at those locations.  However, while this
scenario would explain the location of at least some of the MIR peaks,
it would also create a new problem: why should the gas inflow have
stopped 5.6\,Myr ago, at least to a level where no more massive
young clusters are being formed?  Since we have no additional evidence
for such a scenario we don't consider it very likely.

In summary, the central cluster formation in NGC\,7552 is not dominated
by a simple bar/ring geometry, but it is not completely decoupled from
the ring motion either.  Either some negative feedback effects are at
play \citep{dale05} or the gas density will be built up more gradually
over a larger ring segment, and the local physical conditions become
dominant.


\section{Summary}
\label{summary}

We have observed the starburst ring galaxy NGC\,7552 with two
instruments on ESO's VLT, namely the Imager and Spectrograph for the
mid-IR (VISIR) in both imaging and spectroscopy mode, and the
Spectrograph for INtegral Field Observations in the Near Infrared
(SINFONI) at the VLT.  The angular resolution of the VISIR
observations is 0\farcs$3 - 0$\farcs4, very close to the diffraction
limit of VLT at 10\m.  We compared these data to ISO and Spitzer
observations. Within the given uncertainties, the agreement between
the measured fluxes from the different instruments and observing modes
is very good.  Our main findings can be summarized as follows:

\begin{itemize}
\item{The starburst ring is clearly identifiable at MIR wavelengths.
  The \NeII12.8\m\ line emission correlates well with the NIR
  \brg\ emssion.}  
\item{We find an excellent correspondence between the \NeII\ emission
  regions and the highly dust extincted dark ring in optical HST
  observations.  The latter look very similar to the K band
  continuum.}
\item{We have identified nine peaks of unresolved MIR emission.  Our
  strongest 12\m\ source does not show significant emission in the
  K~band continuum.}
\item{We do not detect MIR emission from the nucleus of NGC\,7552,
  which is very prominent at optical and NIR continuum wavelengths.}
\item{The multi-wavelength data indicate that the commonly used term
  `starburst ring' is insufficient to properly characterize a rather
  complex picture of cluster formation within an annular region of
  more than 100 parsec width.}
\item{A large fraction of the \NeII\ and PAH band emission is not
  directly associated with the location of the massive, young star
  clusters.  At the cluster positions we measured about 32\% of the
  total \NeII\ emission of the ring but only 5\% of the PAH emission.}
\item{A comparison of the strength of the 11.3\m\ PAH feature
  between VISIR and \spitzer\ spectra indicates that zooming in on the
  clusters on scales of 33\,pc reduces the observed PAH strength by a
  factor of six.  This can be explained by spatially resolving the
  \HII\ region from the PDR complex.}  
\item{The individual cluster properties, in terms of infrared
  luminosities, ages, Lyman continuum photons, and extinction, are in
  good agreement with the properties of other MIR-selected massive
  clusters in other galaxies.}
\item{From the ratio of \brg/Pa$\beta$ we computed an extinction map.
  The average extinction of the nine peaks is $A_{\rm V} = 7.4$ and
  reaches from 5.2 to 9.0.  While the average extinction of our
  MIR-selected clusters is more than five times higher than that of
  K~band selected clusters in the Antennae, it is similar to
  MIR-selected clusters in other dense starburst environments.
  Apparently, the mechanism of gas and dust removal is less efficent
  in these extreme environments.}
\item{The total luminosity of the nine MIR peak is $L_{IR} = 2.1\times
  10^{10}$\lsun, which is about 75\% of the bolometric luminosity of
  the starburst ring.  In the continuous star formation approximation
  this would correspond to a SFR of 3.7\msun\,yr$^{-1}$.}
\item{We determined the cluster ages from the equivalent width of the
  \brg\ line. They lie within the range of $5.9\pm0.3$~Myr.
  Independently, we estimated a ratio of \NeIII\,/\,\NeII\,$\approx
  0.41$, which is in good agreement with an age of approximately
  5~Myr.}
\item{The youngest massive clusters are located near the Northern
  contact point of the ring while the oldeer peaks are observed
  further away from the contact points.  However, the age spread among
  the clusters of 0.8\,Myr is small compared to the travel time of
  $\sim5.6$\,Myr for half an orbit within the starburst ring.  We
  cannot rule out that the clusters were formed near the contact
  points more than 5\,Myr ago, but we find no strong evidence for the
  scenario where the continuous inflow of gas leads to the ongoing
  formation of massive clusters near the contact points. It appears
  more likely that the gas density build up more gradually over larger
  ring segments, and that the local physical conditions then determine
  the cluster formation.}
\item{Even at the resolution of VISIR on an eight-meter VLT, we cannot
  rule out that an unresolved MIR peak is not a single super star
  cluster but may consist of several sub-clusters, possibly with some
  age spread.  If so, this would affect the derived ages, the slope of
  the IMF, and dynamical cluster mass estimates.  This issue can only
  be solved with milli-arcsecond resolution at MIR wavelengths as will
  be provided by the next generation of extremely large telescopes
  (ELTs).}
\end{itemize}

\begin{acknowledgements}
      We would like to thank Brent Groves for helpful discussions
      concerning the cluster age estimates, and Brad Whitmore for a
      discussion on the observational aspects of dust extinguished
      clusters.  We would also like to thank the anonymous referee for
      spotting an error in the NIR data reduction in the draft
      version.
\end{acknowledgements}

\bibliographystyle{aa} 
\bibliography{biblio} 

\end{document}